\documentclass[aps,prb, onecolumn, amssymb,floatfix,nofootinbib, showkeys, showpacs, superscriptaddress]{revtex4-2}
\usepackage{amsmath}
\usepackage{tabularx}
\usepackage{bm}
\usepackage{slashed}
\usepackage{euscript}
\usepackage{graphicx}
\usepackage{color}
\usepackage{amsfonts}
\usepackage{exscale}
\usepackage{amsbsy}
\usepackage{subfigure}
\usepackage{textcomp}
\usepackage{comment}
\usepackage{hyperref}
\usepackage{lineno}
\usepackage{booktabs}

\usepackage{setspace}
\begin{document}

\title{Phonon-mediated spin transport in quantum paraelectric metals}
\date{\today}
\author{Kyoung-Min Kim}
\email{kmkim@ibs.re.kr}
\affiliation{Center for Theoretical Physics of Complex Systems, Institute for Basic Science, Daejeon 34126, Republic of Korea}

\author{Suk Bum Chung}
\email{sbchung0@uos.ac.kr}
\affiliation{Department of Physics and Natural Science Research Institute,
University of Seoul, Seoul 02504, Republic of Korea}
\affiliation{School of Physics, Korea Institute for Advanced Study, Seoul 02455, Republic of Korea}

\begin{abstract}
The concept of ferroelectricity is now often extended to include continuous inversion symmetry-breaking transitions in various metals and doped semiconductors. Paraelectric metals near ferroelectric quantum criticality, which we term `quantum paraelectric metals,' 
possess soft transverse optical phonons 
which can have Rashba-type coupling to itinerant electrons in the presence of spin-orbit coupling. We find through the Kubo formula calculation that such Rashba electron-phonon coupling has a profound impact on electron spin transport. While the spin Hall effect arising from non-trivial electronic band structures has been studied extensively, we find here the presence of the Rashba electron-phonon coupling can give rise to spin current, including spin Hall current, in response to an inhomogeneous electric field even with a completely trivial band structure. Furthermore, this spin conductivity displays unconventional characteristics, such as quadrupolar symmetry associated with the wave vector of the electric field and a thermal activation behavior characterized by scaling laws dependent on the phonon frequency to temperature ratio. These findings shed light on exotic electronic transport phenomena originating from ferroelectric quantum criticality, highlighting the intricate interplay of charge and spin degrees of freedom.
\end{abstract}

\keywords{spin  conductivity, ferroelectric quantum criticality, Rashba spin-orbit coupling, transverse optical phonons, Kubo formula}

\pacs{72.10.Di, 72.25.Bq, 77.90.+k}

\maketitle
\pagebreak

\section{Introduction}

\begin{figure} [ht]
    \centering
    \includegraphics[width=.4\textwidth]{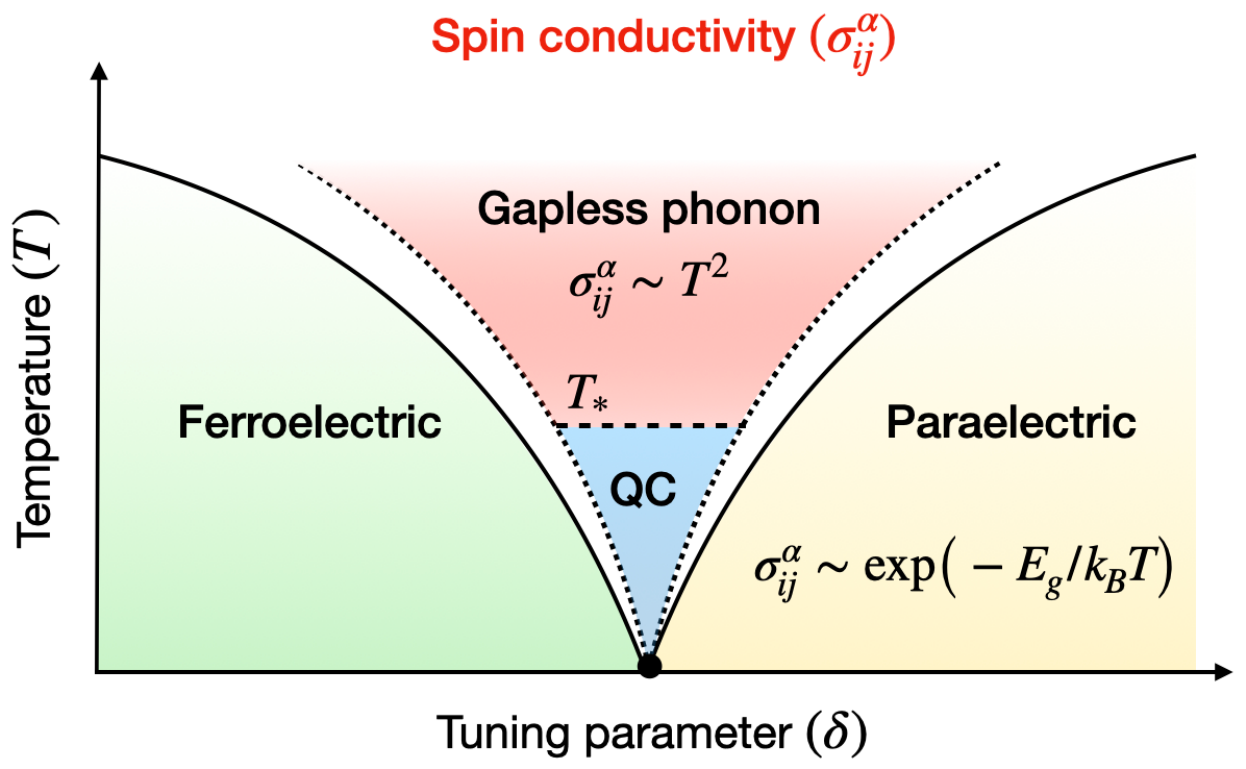}
    \caption[Phase diagram]{\textbf{Spin conductivity in quantum paraelectric metals.} Schematic phase diagram of quantum paraelectric metals with spin-orbit coupling near the ferroelectric quantum critical point $(\delta, T)=(0, 0)$ ($\delta$: tuning parameter, $T$: temperature) and distinct scaling laws of spin conductivity ($\sigma_{ij}^\alpha$) in each phase, where $\alpha,i,j$ denote the direction of spin quantization axis, spin current and electric fields, respectively. In these quantum paraelectric metals, an inhomogeneous electric field induces a spin current. In the paraelectric phase (yellow area), the phonon-mediated spin conductivity possesses a thermal-activated form: $\sigma_{ij}^\alpha \sim \exp{(-E_g/k_BT)}$, where $E_g\sim \delta$ is the energy gap of soft transverse optical (TO) phonons associated with the phase transition. In contrast, within the gapless-phonon region (red area), the spin conductivity adopts a power-law behavior: $\sigma_{ij}^\alpha \sim T^2$. This power law of spin conductivity may undergo modifications in the quantum critical (QC) region (blue area) below a specific crossover temperature scale $T_*$ (indicated by the dashed line), wherein phonons or electrons lose their coherence due to significant self-energy effects. $\delta$ and $T$ are, respectively, the tuning parameter and temperature, while  $\alpha$, $i$, $j$ denote the direction of spin quantization axis, spin current, and electric field, respectively.} 
    \label{fig1}
\end{figure}

The search for materials combining ferroelectricity/polarity with metallicity has been longstanding in condensed matter physics, dating back to the first proposal by Anderson and Blount over 50 years ago \cite{PhysRevLett.14.217}. This endeavor has made significant progress, especially in the last decade, resulting in the accumulation of numerous experimentally confirmed examples \cite{Zhou_2020, PhysRevMaterials.7.010301, Bhowal2023, Hassan2020}, starting with LiOsO$_3$ \cite{shi2013ferroelectric}. Other noteworthy examples include doped quantum paraelectrics such as SrTiO\textsubscript{3} \cite{annurev:/content/journals/10.1146/annurev-conmatphys-031218-013144, Rischau2017, PhysRevLett.125.087601, Salmani-Rezaie2020}, IV-VI compounds \cite{Yu2018, Bilz1983}, and 
certain transition metal dichalcogenides \cite{doi:10.1126/sciadv.1601378, Fei2018, Yang2018,  doi:10.1126/sciadv.aax5080}. These so-called ferroelectric (or polar) metals, typically doped ferroelectrics in semimetals and semiconductors, present the intriguing coexistence of ferroelectricity and metallicity, contrary to their apparent mutual exclusivity. In addition, the possibility of various correlated electronic phenomena arising from ferroelectric quantum fluctuations near a ferroelectric quantum critical point, including the augmentation of the critical temperature for superconductivity, has attracted strong interest \cite{annurev:/content/journals/10.1146/annurev-conmatphys-031218-013144, Rowley2014, PhysRevLett.115.247002, Rischau2017, PhysRevX.9.031046, Volkov2022, GASTIASORO2020168107, Chandra_2017, Yu2022, PhysRevB.104.L220506, PhysRevB.106.L121114}.

For the displacive ferroelectrics under consideration, the continuous ferroelectric phase transition involves the softening of transverse optical (TO) phonon modes associated with the displacement in proximity to the critical point \cite{Cohen1992, PhysRevLett.109.247601}, as this transition is characterized by a collective displacement of ions from their centrosymmetric positions \cite{PhysRevLett.125.087601, Chandra_2017, 10.1093/acprof:oso/9780198507789.001.0001}. Given that the TO mode displacement breaks the inversion symmetry while preserving the time-reversal symmetry, the interactions between the TO phonons and itinerant electrons in the presence of any finite atomic spin-orbit coupling takes the unconventional form of a Rashba-type spin-orbit coupling, which couples the momentum and spin of itinerant electrons \cite{PhysRevB.99.094524, MLee2020, PhysRevB.101.174501, PhysRevB.107.165110}. We refer to these distinctive interactions as “phonon-mediated spin-orbit coupling” (PM-SOC). Previous theoretical studies explored the impacts of the PM-SOC on correlated electronic phenomena in the quantum critical region, such as non-Fermi liquid behavior \cite{PhysRevB.107.165110}
, enhanced superconducting instability \cite{PhysRevB.107.165110, Yu2022}, 
charge transport \cite{PhysRevB.107.165110} 
and optical conductivity \cite{ PhysRevB.105.125142}; transport effects of soft TO phonons 
have also been investigated for 
the two-phonon scattering mechanism \cite{PhysRevB.104.115201, PhysRevB.106.L121114}. However, the effect of the PM-SOC on 
spin transport, particularly when subject to inhomogeneous electric fields, remains unexplored so far, remaining a missing piece of the physics near the ferroelectric quantum critical region.

In this study, we investigate the influence of the PM-SOC on the electronic transport properties of a centrosymmetric metal ({\it i.e.} possessing finite carrier concentration) near the ferroelectric quantum critical point, which may be termed as a `quantum paraelectric metal.' From the Kubo formula, we obtain a nonzero spin conductivity, even in the centrosymmetric paraelectric phase (Fig.~\ref{fig1}), from a single orbital, albeit contingent on the presence of inhomogeneous external electric fields. This phenomenon may seem counter-intuitive at first glance since not only is the Rashba spin-orbit coupling in the electronic band structure, which results in the finite spin Hall conductivity \cite{PhysRevLett.92.126603, SQShen2004}, symmetry-forbidden in such a phase, but any orbital Hall effect \cite{TTanaka2008, Kontani2009, SRPark2011, Go2018} is also absent. However, the Rashba-type spin-orbit coupling to TO phonons \cite{PhysRevLett.127.237601}, {\it i.e.}, the PM-SOC, in conjunction with inhomogeneous external electric fields, gives rise to an unconventional type of spin conductivity. Notably, we shall show in Sec.~\ref{sec:quad} that this phonon-mediated spin conductivity exhibits a unique directional dependence on the wave vector of external electric fields, displaying a quadrupolar symmetry with respect to the wave vector that is, however, distinct from the quadrupolar symmetry predicted for electrical Hall resistivity in quantum Hall states \cite{haldane2009hall, PhysRevLett.108.066805} or spin Hall conductivity in Rashba metals \cite{Zhang2022}. Furthermore, we demonstrate in Sec.~\ref{sec:thermal} that our phonon-mediated spin conductivity also exhibits peculiar scaling laws as a function of a tuning parameter and temperature (Fig.~\ref{fig1}). Whereas most theoretical research on spin transport has been based on band structure considerations, our findings point to new possibilities in interaction-induced spin transport. Moreover, our results highlight the intriguing aspect of the emergent exotic transport phenomena arising from the intricate interplay of charge and spin degrees of freedom in itinerant electrons in the realm of ferroelectric quantum criticality.

\section{Results}

\subsection{Model}

Quantum paraelectric metals are characterized by the emergence of soft TO phonon modes and their distinctive electron-phonon interactions, which, in combination with atomic spin-orbit coupling, exhibit a Rashba-type spin-orbit coupling for itinerant electrons \cite{SRPark2011, GASTIASORO2020168107, MLee2020}. The minimal model for a quantum paraelectric metal is given by the following effective Hamiltonian \cite{PhysRevB.101.174501, PhysRevB.107.165110}:
\begin{eqnarray} \label{eq:H}
    \hat{H}=\sum_{s,\bm{p}}\xi_{\bm{p}}c_{s,\bm{p}}^\dagger c_{s,\bm{p}} + \frac{g}{\sqrt{V}} \sum_{\bm{k}}\sum_{s,s',\bm{p}} \bm{\phi}_{\bm{k}}\cdot(\bm{\sigma}_{ss'}\times\bm{p})c_{s,\bm{p}+\bm{k}/2}^\dagger c_{s',\bm{p}-\bm{k}/2}
\end{eqnarray}
for electrons, where $c_{s,\bm{p}}^\dagger$ is the electron creation operator ($s$, $\bm{p}$ denoting the spin and wave vector of the electron, respectively), the electron energy dispersion is isotropic, {\it i.e.} $\xi_{\bm{p}}=\frac{\hbar^2|\bm{p}|^2}{2m}-\mu$ ($m$ and $\mu$ denoting the electron effective mass and the chemical potential, respectively), $g$ is the coupling constant for the electron-phonon interaction, and $\bm{\phi}_{\bm{k}}$ is the transverse phonon displacement field, whose dynamics, in the free limit, is given by the action
\begin{equation}
    S_{ph}= \frac{M_{ph}}{\hbar^2}\sum_{i\nu_n}\sum_{\bm{k}}\sum_{ij} \phi_i (-i\nu_m,-\bm{k})(\nu_n^2 + \omega^2_{\bm{k}})\mathcal{P}_{ij}(\bm{k})\phi_j(i\nu_m,\bm{k}), 
\end{equation} 
where $\mathcal{P}_{ij}(\bm{k})\equiv\delta_{ij}-(\hat{\bm{e}}_i\cdot\hat{\bm{k}})(\hat{\bm{e}}_j\cdot\hat{\bm{k}})$ is the transverse projection operator, $\nu_n$ is the boson Matsubara frequency, $M_{ph}$ is the phonon effective mass, and the phonon energy dispersion is given by $\omega_{\bm{k}}^2=(\hbar c|\bm{k}|)^2+E_g^2$, where $c$ and $E_g$ denote the phonon velocity and energy gap, respectively.

\begin{figure} [t!]
    \centering
    \includegraphics[width=.5\textwidth]{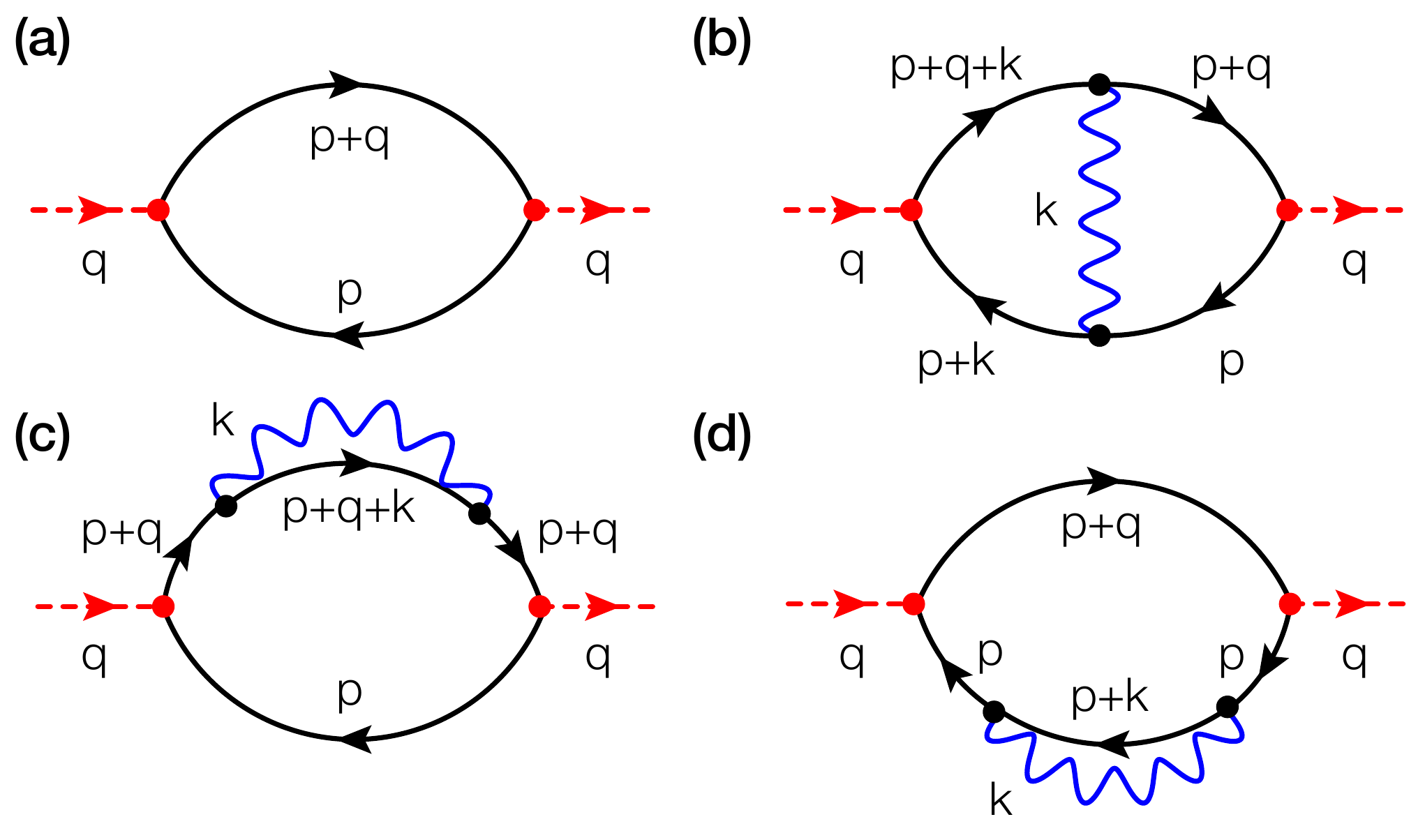}
	\caption{\textbf{Relevant Feynman diagrams for the computation of phonon-mediated spin conductivity.} \textbf{a} The correlation function involving charge and spin currents at the tree level. 
\textbf{b}–\textbf{d} The correlation functions at the one-loop order. The black solid and blue wavy lines represent electron and phonon propagators, respectively. The black vertex points denote the electron-phonon interactions that result in Rashba-type spin-orbit coupling for electrons. The red lines represent external electric fields. In each diagram, $\bm{p}$, $\bm{k}$, and $\bm{q}$ represent the wave vectors for electrons, phonons, and external electric fields, respectively.}
\label{fig2}
\end{figure}

Whereas previous studies of the Eq.~\eqref{eq:H} quantum paraelectric metal model focused on its instability to superconductivity \cite{PhysRevB.101.174501, PhysRevB.107.165110} or optical conductivity \cite{PhysRevB.105.125142}, we calculate in this work its spin conductivity, denoted as $\sigma_{ij}^\alpha(\omega,\bm{q})$, using the Kubo formula:
\begin{align} \label{eq:Kubo}
    \sigma_{ij}^\alpha(\omega,\bm{q})=\frac{i}{\omega}\pi_{ij}^\alpha(i\omega_n,\bm{q})\big\vert_{i\omega_n\rightarrow\omega+i\delta}.
\end{align}
Here, the indices $i$ and $j$ denote the directions of the spin and charge currents, respectively, while $\omega$ and $\mathbf{q}$ denote the frequency and wave vector of the external electric field. $\pi_{ij}^\alpha(i\omega_n,\bm{q})$ represents the current-current correlation function, defined as
\begin{align} \label{eq:pi}
    \pi_{ij}^\alpha(i\omega_n,\bm{q})=&\;-\frac{1}{V}\int^\beta_0d\tau e^{i\omega_n\tau}\big\langle j_{\textrm{spin},i}^\alpha(\tau, \bm{q})j_j(0,-\bm{q})\big\rangle,
\end{align}
where $j_{\textrm{spin},i}^\alpha(\bm{q})$ and $j_j(\bm{q})$ denote the spin and charge current operators, respectively, and $\omega_n$ the fermion Matsubara frequency. The average $\langle \cdots\rangle$ denotes the ensemble average over the quantum partition function. The charge and spin current operators are explicitly given by
\begin{align}
    j_{\textrm{spin},i}^{\alpha}(\bm{q})=&\;\frac{\hbar^2}{m}\sum_{s,s',\bm{p}}\bigg[\hat{\bm{e}}_i\cdot\bigg(\bm{p}+\frac{1}{2}\bm{q}\bigg)\bigg](\sigma_\alpha)_{ss'}c_{s,\bm{p}+\bm{q}}^\dagger c_{s',\bm{p}},\\
    j_j(\bm{q})=&\;\frac{e\hbar}{m}\sum_{s,\bm{p}}\bigg[\hat{\bm{e}}_j\cdot\bigg(\bm{p}+\frac{1}{2}\bm{q}\bigg)\bigg]c_{s,\bm{p}+\bm{q}}^\dagger c_{s,\bm{p}}.
\end{align}
Here, the unit vectors $\hat{\bm{e}}_i$ and $\hat{\bm{e}}_j$ denote the directions of the spin and charge currents, respectively. We compute $\sigma_{ij}^\alpha(\omega,\bm{q})$ through a diagrammatic expansion in $g$. Further details can be found in Supplementary Note 1. As a result, we obtain the following expression for $\sigma^\alpha_{ij}(\omega,\bm{q})$ \cite{mahan}:
\begin{align} \label{eq:sigma_1}
    \sigma^\alpha_{ij}(\omega,\bm{q}) = & \; \frac{e\hbar^4}{m^2}\int\frac{d^3p}{(2\pi)^3}\hat{\bm{e}}_j\cdot\bigg(\bm{p}-\frac{1}{2}\bm{q}\bigg)\int^\infty_{-\infty}\frac{d\epsilon}{2\pi}\frac{1}{\omega}\bigg[n_F(\epsilon)-n_F(\epsilon+\omega)\bigg]\nonumber\\
    & \times G_\textrm{ret}(\epsilon+\omega,\bm{p}+\bm{q}) G_\textrm{adv}(\epsilon,\bm{p}) \gamma_i^\alpha(\epsilon+\omega+i\delta, \epsilon-i\delta;\bm{p}+\bm{q}, \bm{p}). 
\end{align}
Here, $n_F(\epsilon)$ denotes the Fermi-Dirac distribution function and $G_\textrm{ret}(\epsilon+\omega,\bm{p}+\bm{q})$, $G_\textrm{adv}(\epsilon,\bm{p})$ the 
retarded and advanced propagators for the electron and hole states, respectively. We focus on the paraelectric phase just outside the quantum critical region. 
In this case, we may posit that both electron and phonon propagators possess well-defined quasiparticle peaks: $G_\textrm{ret}(\epsilon+\omega,\bm{p}+\bm{q})=\frac{1}{\epsilon+\omega-\xi_{\bm{p}+\bm{q}}+i\hbar/2\tau}$, $G_\textrm{adv}(\epsilon,\bm{p})=\frac{1}{\epsilon-\xi_{\bm{p}}-i\hbar/2\tau}$, and $D_{ij}(\omega,\bm{k})=\mathcal{P}_{ij}(\bm{k})\frac{2\omega_{\bm{k}}}{(\omega+i\delta)-(\omega_{\bm{k}})^2}$. Here, $\tau$ represents the quasiparticle lifetime stemming from elastic disorder scattering present in realistic macroscopic materials 
and is phenomenologically introduced without explicit modeling for the relevant disorder scattering.

\subsection{Phonon-mediated spin conductivity}

\begin{figure}[tb!]
    \centering   
    \includegraphics[width=.5\textwidth]{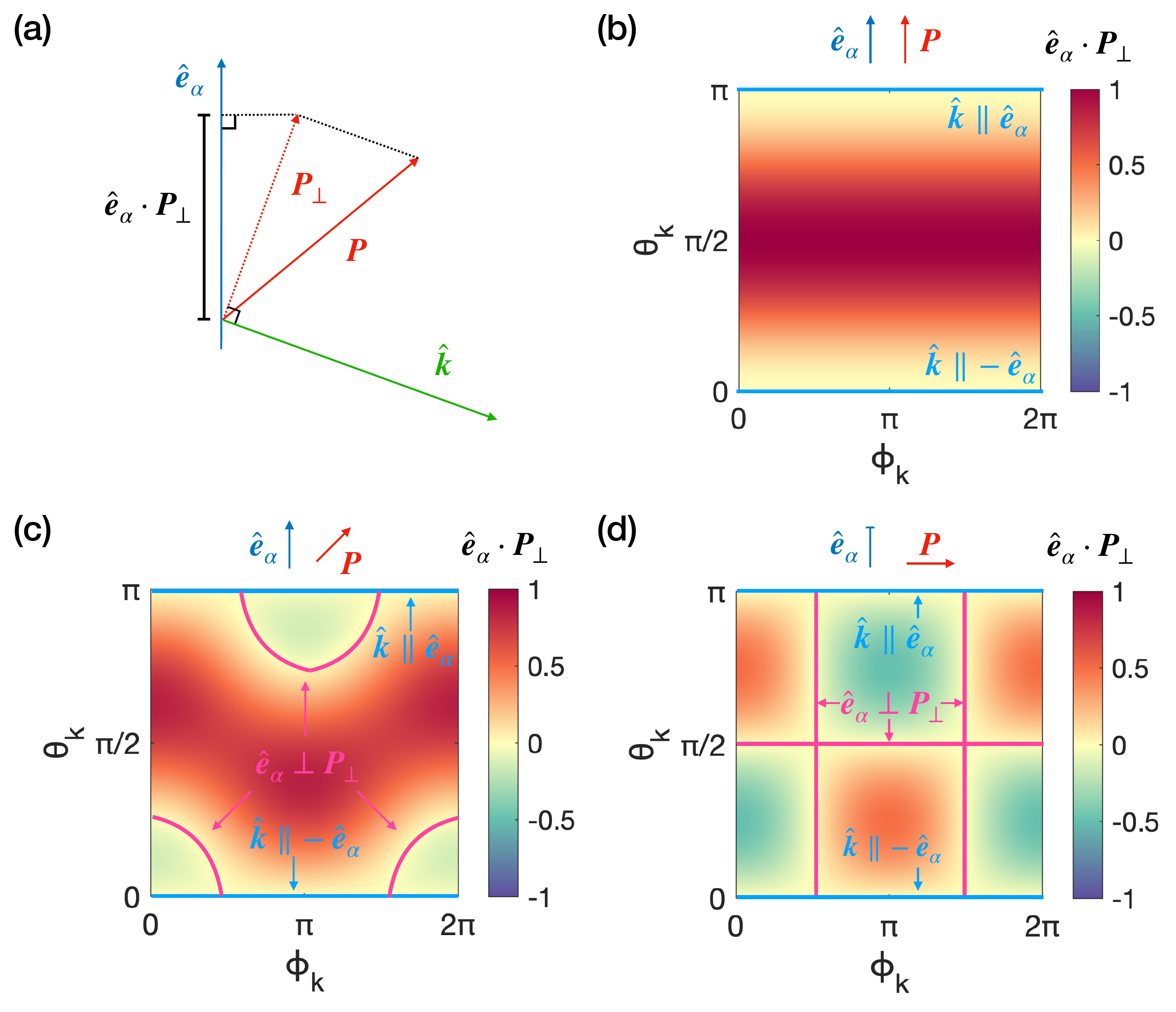}
	\caption{\textbf{Phonon-mediated spin polarization.} \textbf{a} Schematic illustration of the vector calculations involved in computing phonon-mediated spin polarization. The outcome, $\hat{\bm{e}}_\alpha\cdot\bm{P}_\perp$, represents the value of the desired spin polarization. (b-d) Plots of $\hat{\bm{e}}_\alpha\cdot\bm{P}_\perp$ as a function of the azimuthal ($\phi_k$) and polar ($\theta_k$) angles of $\hat{\bm{k}}=(\sin{\theta_k}\cos{\phi_k},\sin{\theta_k}\sin{\phi_k},\cos{\theta_k})$. \textbf{b} Plot for $\bm{P}\parallel\hat{\bm{e}}_\alpha$, where $\hat{\bm{e}}_\alpha=\hat{z}$ and $\bm{P}=|\bm{P}|\hat{z}$. \textbf{c} Plot for $\hat{\bm{e}}_\alpha=\hat{z}$ and $\bm{P}=|\bm{P}|(\sqrt{2}/2,0,\sqrt{2}/2)$. \textbf{d} Plot for $\bm{P}\perp\hat{\bm{e}}_\alpha$, where $\hat{\bm{e}}_\alpha=\hat{z}$ and $\bm{P}=|\bm{P}|\hat{x}$. The unit vectors $\hat{\bm{e}}_\alpha$ and $\hat{\bm{k}}$ denote the directions of the spin polarization and the phonon wave vector, respectively. $\bm{P}=\bm{q}\times(\bm{p}+\bm{k}/2)$ denotes the vector product of $\bm{q}$ and $\bm{p}+\bm{k}/2$. $\bm{P}_\perp = \bm{P} - \hat{\bm{k}} (\hat{\bm{k}}\cdot\bm{P})$ represents the vector rejection of $\bm{P}$ from $\hat{\bm{k}}$. \textbf{b}-\textbf{d} The color scale denotes the value of $\hat{\bm{e}}_\alpha\cdot\bm{P}_\perp$. Specific relative orientations between $\hat{\bm{e}}_\alpha$ and $\bm{P}$ are utilized in each plot. The cyan and magenta lines indicate the conditions under which $\hat{\bm{e}}_\alpha\cdot\bm{P}_\perp$ vanishes, corresponding to $\hat{\bm{k}}||\pm\hat{\bm{e}}_\alpha$ and $\hat{\bm{e}}_\alpha\perp\bm{P}_\perp$, respectively.}
    \label{fig3}
\end{figure}

From the yet unelucidated vertex term $\gamma^\alpha_i$, we can show that the spin conductivity of Eq.~\eqref{eq:sigma_1} vanishes in lieu of the virtual TO phonon exchange. At the tree level as shown in Fig.~\ref{fig2}a, where such exchange is absent, it arises entirely from $j_{\textrm{spin},i}^{\alpha}(\bm{q})$ and takes the form of $\gamma_i^\alpha(\epsilon-i\delta,\epsilon+\omega+i\delta;\bm{p},\bm{p}+\bm{q})=\hat{\bm{e}}_i\cdot(\bm{p}+\frac{1}{2}\bm{q})\textrm{tr}(\sigma_\alpha)$, where “tr” represents a trace over spin matrices, and hence vanishes. At the one-loop order, there are three diagrams, as shown by Fig.~\ref{fig2}b–d. We note that two of these, Fig.~\ref{fig2}c–d (the “self-energy diagrams”), vanish and hence do not contribute to the PMSP 
due to the Pauli matrix algebra (see 
Supplementary Note 3 for details). 
However, the “vertex-correction diagram” of Fig.~\ref{fig2}b allows for a non-vanishing form of the vertex function through the PM-SOC in this system:
\begin{align} \label{eq:gamma} 
    \gamma_i^\alpha(\epsilon-i\delta,\epsilon+\omega+i\delta;\bm{p},\bm{p}+\bm{q})=&\; \frac{g^2\hbar^2}{M_{ph}} \int\frac{d^3k}{(2\pi)^3}\frac{1}{2\omega_{\bm{k}}} \hat{\bm{e}}_i\cdot\bigg(\bm{p}+\bm{k}+\frac{1}{2}\bm{q}\bigg) \tilde{\gamma}^\alpha(\bm{p},\bm{k};\bm{q}) \mathcal{S}(\epsilon,\bm{p},\bm{k};\omega,\bm{q}).
\end{align}
Here, $\mathcal{S}(\epsilon,\bm{p},\bm{k};\omega,\bm{q})$ accounts for the creation of a virtual phonon with a wave vector $\bm{k}$, leading to the transition of the electron and hole states into $(\bm{p}+\bm{q}+\bm{k})$ and $(\bm{p}+\bm{k})$, respectively. Such a factor is a common occurrence in conventional electron-phonon systems \cite{mahan}. To derive this factor, we perform a Matsubara frequency summation in $\gamma_i^\alpha(\epsilon-i\delta,\epsilon+\omega+i\delta;\bm{p},\bm{p}+\bm{q})$ using a contour integral, following standard transport theory calculations \cite{mahan} (see Supplementary Note 2 for details). $\mathcal{S}(\epsilon,\bm{p},\bm{k};\omega,\bm{q})$ is explicitly given by
\begin{align} \label{eq:S_main}
    \mathcal{S}(\epsilon,\bm{p},\bm{k};\omega,\bm{q}) = & \; \bigg[n_B(\omega_{\bm{k}})+\frac{1}{2}n_F(\epsilon+\omega_{\bm{k}})+\frac{1}{2}n_F(\epsilon+\omega+\omega_{\bm{k}})\bigg]G_{\textrm{adv}}(\epsilon+\omega_{\bm{k}},\bm{p}+\bm{k})G_{\textrm{ret}}(\epsilon+\omega+\omega_{\bm{k}},\bm{p}+\bm{k}+\bm{q})\nonumber\\
    & + \bigg[n_B(\omega_{\bm{k}})+1-\frac{1}{2}n_F(\epsilon-\omega_{\bm{k}})-\frac{1}{2}n_F(\epsilon+\omega-\omega_{\bm{k}})\bigg]G_{\textrm{adv}}(\epsilon-\omega_{\bm{k}},\bm{p}+\bm{k})G_{\textrm{ret}}(\epsilon+\omega-\omega_{\bm{k}},\bm{p}+\bm{k}+\bm{q}).
\end{align}
$\tilde{\gamma}^\alpha(\bm{p},\bm{k};\bm{q})$ now includes the effects of the PM-SOC vertex pair as shown in Fig.~\ref{fig2}b:
\begin{align} \label{eq:gamma_tilde_1}
    \tilde{\gamma}^\alpha(\bm{p},\bm{k};\bm{q}) = \sum_{k,l}\mathcal{P}_{kl}(\bm{k})\textrm{tr}\Big[\sigma_\alpha  \hat{\bm{e}}_k \cdot (\bm{\sigma} \times (\bm{p}+\bm{q}+\bm{k}/2)) \hat{\bm{e}}_l \cdot (\bm{\sigma} \times (\bm{p}+\bm{k}/2)) \Big].
\end{align}
Parts of this trace arise from the annihilation and the creation of a virtual TO phonon through the Rashba-type spin-orbit interactions, which result in $\hat{\bm{e}}_k\cdot(\bm{\sigma}\times(\bm{p}+\bm{q}+\bm{k}/2))$ and $\hat{\bm{e}}_l\cdot(\bm{\sigma}\times(\bm{p}+\bm{k}/2))$ respectively. Their product can be written as:  
\begin{align}
    & \hat{\bm{e}}_k \cdot (\bm{\sigma} \times (\bm{p}+\bm{q}+\bm{k}/2)) \hat{\bm{e}}_l \cdot (\bm{\sigma} \times (\bm{p}+\bm{k}/2)) \nonumber \\
    = & \; ((\bm{p}+\bm{q}+\bm{k}/2) \cdot (\bm{p}+\bm{k}/2)) (\hat{\bm{e}}_k \cdot \hat{\bm{e}}_l) - ((\bm{p}+\bm{q}+\bm{k}/2) \cdot \hat{\bm{e}}_l) ((\bm{p}+\bm{k}/2) \cdot \hat{\bm{e}}_k) \nonumber \\
    & + i (\bm{p}+\bm{q}+\bm{k}/2) \cdot(\hat{\bm{e}}_k \times \hat{\bm{e}}_l) ((\bm{p}+\bm{k}/2) \cdot \bm{\sigma}) + i \hat{\bm{e}}_k \cdot (\bm{q} \times (\bm{p}+\bm{k}/2)) (\hat{\bm{e}}_l \cdot \bm{\sigma}).
\end{align}
In this expression, the first two terms vanish after taking the trace over spin due to $\sigma_\alpha$. The third term vanishes after averaging over all phonon polarizations ($k$ and $l$) due to antisymmetry in $k \leftrightarrow l$, as the transverse projection operator $\mathcal{P}_{kl}(\bm{k})$ is symmetric. However, the last term may survive in both tracing over spin and summing over the phonon polarizations, leaving a nonzero spin polarization, 
\begin{align}
    \tilde{\gamma}^\alpha(\bm{p},\bm{k};\bm{q}) = 2i \Big\{\hat{\bm{e}}_\alpha\cdot\big[\bm{q}\times(\bm{p} + \bm{k}/2)\big] - (\hat{\bm{e}}_\alpha\cdot\hat{\bm{k}}) \hat{\bm{k}}\cdot \big[\bm{q}\times(\bm{p}+\bm{k}/2)\big] \Big\}, \label{eq:gamma_tilde_2}
\end{align}
in contrast to the tree level, where $\hat{\bm{e}}_\alpha$ denotes the unit vector in the direction of $\alpha$. This quantity is termed “phonon-mediated spin polarization” (PMSP).

A 
physical picture of 
PMSP can be obtained through a simple re-writing of our PMSP. The first step is to take the vector product of the two momenta associated with the electron and hole states, namely $(\bm{p}+\bm{q}+\bm{k}/2)$ and $(\bm{p}+\bm{k}/2)$: $\bm{P}=\bm{q}\times(\bm{p}+\bm{k}/2)$. The next step is to 
take the $\hat{\bm{k}}$ rejection of $\bm{P}$, {\it i.e.} 
$\bm{P}_\perp = \bm{P} - \hat{\bm{k}} (\hat{\bm{k}}\cdot\bm{P})$, as illustrated in Fig.~\ref{fig3}a. The final step is to take the scalar product of $\bm{P}_\perp$ and $\hat{\bm{e}}_\alpha$, resulting in $\tilde{\gamma}^\alpha(\bm{p},\bm{k};\bm{q})=2i\hat{\bm{e}}_\alpha\cdot\bm{P}_\perp$. From $\hat{\bm{e}}_\alpha\cdot\bm{P}=\bm{q}\cdot[(\bm{p}+\bm{k}/2)\times\bm{\sigma}]$, we can regard $\tilde{\gamma}^\alpha(\bm{p},\bm{k};\bm{q})$ as a Rashba-like term with $\bm{q}$, the wave vector of the external electric field, plays a role analogous to inversion-symmetry breaking fields in conventional Rashba spin-orbit coupling. This in turn induces a nonvanishing spin polarization proportional to $\bm{q}$, namely the PMSP. Also, we can easily  see now conditions under which $\tilde{\gamma}^\alpha(\bm{p},\bm{k};\bm{q})$ would vanish. A simplest example is determined by the
orientations of $\hat{\bm{k}}$:
\begin{align}
    \hat{\bm{k}}\parallel \pm \hat{\bm{e}}_\alpha \Rightarrow \tilde{\gamma}^\alpha(\bm{p},\bm{k};\bm{q})=0,
\end{align}
as depicted by the cyan lines in Fig.~\ref{fig3}b–d. In other words, TO phonons with momenta parallel to $\hat{\bm{e}}_\alpha$ do not contribute to the PMSP and, ultimately, to spin conductivity. Additionally, there exists a more general vanishing condition:
\begin{align}
    \hat{\bm{e}}_\alpha \perp \bm{P}_\perp \Rightarrow \tilde{\gamma}^\alpha(\bm{p},\bm{k};\bm{q})=0,
\end{align}
as illustrated by the magenta lines in Fig.~\ref{fig3}b–d. Consequently, TO phonons that result in $\bm{P}_\perp$ being perpendicular to $\hat{\bm{e}}_\alpha$ also do not contribute to the generation of the PMSP. 

From the PMSP we have obtained, we find that the lowest order term in $\bm{q}$ of the DC spin conductivity, {\it i.e.} 
\begin{equation*}
    \sigma^\alpha_{ij}(\bm{q})=\lim_{\omega\rightarrow0}\sigma^\alpha_{ij}(\omega,\bm{q}),
\end{equation*}
is quadratic, which means that the spin current will arise in response to an inhomogeneous electric field. By substituting Eqs. \eqref{eq:gamma}, \eqref{eq:S_main} and \eqref{eq:gamma_tilde_2} into Eq.~\eqref{eq:sigma_1} and taking the limit $\omega\rightarrow0$, we obtain
\begin{align} \label{eq:sigmaDC_1}
    \sigma^{\alpha}_{ij}(\bm{q})=&\;\frac{i}{\pi}\frac{g^2e\hbar^6}{m^2M_{ph}}\int\frac{d^3p}{(2\pi)^3}\int\frac{d^3k}{(2\pi)^3} \frac{n_B(\omega_{\bm{k}})+n_F(\omega_{\bm{k}})}{2\omega_{\bm{k}}} \hat{\bm{e}}_i\cdot\bigg(\bm{p}+\bm{k}+\frac{1}{2}\bm{q}\bigg)\hat{\bm{e}}_j\cdot\bigg(\bm{p}-\frac{1}{2}\bm{q}\bigg) \nonumber \\
    &\times \hat{\bm{e}}_\alpha \cdot \Big\{[\bm{q}\times(\bm{p} + \bm{k}/2)] - \hat{\bm{k}} \big(\hat{\bm{k}} \cdot [\bm{q}\times(\bm{p} + \bm{k}/2)] \big) \Big\} \nonumber \\
    &\times\Big[G_{\textrm{ret}}(0, \bm{p}+\bm{q})G_{\textrm{adv}}(0, \bm{p})G_{\textrm{ret}}(\omega_{\bm{k}},\bm{p}+\bm{k}+\bm{q})G_{\textrm{adv}}(\omega_{\bm{k}},\bm{p}+\bm{k})+(\omega_{\bm{k}} \longleftrightarrow -\omega_{\bm{k}})\Big].
\end{align} 
Given that our PMSP factor is proportional to $|\bm{q}|$, $\sigma^\alpha_{ij}(\bm{q})$ obviously vanishes as $|\bm{q}|\rightarrow0$. However, $\sigma^\alpha_{ij}(\bm{q})$ turns out to be {\it even}, not odd, in $\bm{q}$. Given that the electron dispersion is even in momentum, {\it i.e.} $\xi_{-\bm{p}}=\xi_{\bm{p}}$, we expect the electron and the hole propagators to be even in their momenta. Since the phonon dispersion is likewise even in momentum, {\it i.e.} $\omega_{-\bm{k}}=\omega_{\bm{k}}$, it is straightforward to show that Eq.~\eqref{eq:sigmaDC_1} is unchanged by reversing the sign of all its momenta, {\it i.e.} $\bm{p} \leftrightarrow -\bm{p}$, $\bm{k} \leftrightarrow -\bm{k}$ and $\bm{q} \leftrightarrow -\bm{q}$. Consequently, if we expand $\sigma^\alpha_{ij}(\bm{q})$ in the powers of $\bm{q}$, the first non-vanishing term is quadratic, hence the predominant $\bm{q}$-dependence of $\sigma^\alpha_{ij}(\bm{q})$. Our numerical integration on Eq.~\eqref{eq:sigmaDC_1} confirms this quadratic behavior within a low-$|\bm{q}|$ regime (Fig.~\ref{fig4}b). Assuming $E_F  = 1$ eV and $\tau = 10^{-14}$ s, we determined that the quartic term is significantly smaller than the quadratic term for $|\bm{q}| \ll \frac{\hbar}{E_F\tau } k_F \approx 0.0212 k_F$, where $k_F$ denotes the Fermi wave vector. Further details can be found in Supplementary Note 5.

\begin{table}[t!]
    \centering
    \begin{tabular}{cccc} \toprule
         $i$ & $\alpha$ & $\sigma_{ii}^{\alpha}(\bm{q})$  \\ \midrule
         $y$ & $x$ & $\chi_0 (-2q_z q_y)$ \\ \midrule
         $z$ & $x$ & $\chi_0 (2q_y q_z)$ \\ \midrule
         $x$ & $y$ & $\chi_0 (2q_z q_x)$ \\ \midrule
         $z$ & $y$ & $\chi_0 (-2q_x q_z)$ \\ \midrule
         $x$ & $z$ & $\chi_0 (-2q_y q_x)$ \\ \midrule
         $y$ & $z$ & $\chi_0 (2q_xq_y)$ \\ \bottomrule
    \end{tabular}
    \caption{\textbf{Nonvanishing components of longitudinal conductivity} $\sigma_{ii}^\alpha(\bm{q})$ for $i,\alpha=\{x,y,z\}$ denotes the longitudinal conductivity for spin currents, as defined in Eq.~\eqref{eq:sigmaDC_3}. Here, $i$ denotes both the spin current and the external electric field direction, while $\alpha$ denotes the spin quantization axis direction. $\bm{q}=(q_x,q_y,q_z)$ denotes a wave vector of external electric fields. $\chi_0$ is a $\bm{q}$-independent constant, as defined in Eq.~\eqref{eq:chi0}. } 
    \label{tab1}
\end{table} 

\begin{table}[t!]
    \centering
    \begin{tabular}{cccc}  \toprule
    $i$ & $j$ & $\alpha$ & $ \sigma_{ij}^{\alpha}(\bm{q})  = [\sigma_{ji}^{\alpha}(\bm{q}) ]^*$ \\ \midrule
	$x$&$y$&$x$&$\chi_0(-q_x q_z)+i\kappa_0(-q_x q_z)$ \\ \midrule
         $y$& $z$ & $x$ & $\chi_0(q_y^2 - q_z^2) + i\kappa_0(q_y^2 + q_z^2) $ \\  \midrule
         $z$ & $x$ & $x$ & $\chi_0 (q_x q_y) + i\kappa_0 (-q_x q_y)$ \\ \midrule
         $x$ & $y$ & $y$ & $\chi_0 (q_y q_z) + i\kappa_0 (-q_y q_z)$ \\ \midrule
         $y$ & $z$ & $y$ & $\chi_0 (-q_x q_y) + i\kappa_0 (-q_x q_y)$ \\ \midrule
         $z$ & $x$ & $y$ & $\chi_0 (q_z^2 - q_x^2) + i\kappa_0 (q_z^2 + q_x^2)$ \\ \midrule
         $x$ & $y$ & $z$ & $\chi_0 (q_x^2 - q_y^2) + i\kappa_0 (q_x^2 + q_y^2)$ \\ \midrule
         $y$ & $z$ & $z$ & $\chi_0 (q_xq_z) + i\kappa_0 (-q_xq_z)$ \\ \midrule
         $z$ & $x$ & $z$  & $\chi_0 (-q_y q_z ) + i\kappa_0 (-q_y q_z )$ \\ \bottomrule
    \end{tabular}
    \caption{\textbf{Nonvanishing components of transverse conductivity} $\sigma_{ij}^\alpha(\bm{q})$ for $i,j,\alpha=\{x,y,z\}$ denotes the spin conductivity, as defined in Eq.~\eqref{eq:sigmaDC_3}. Here, $i$ denotes the spin current direction, $j$ denotes the external electric field direction, and $\alpha$ denotes the spin quantization axis direction. $\bm{q}=(q_x,q_y,q_z)$ denotes a wave vector of external electric fields. $\chi_0$ and $\kappa_0$ are $\bm{q}$-independent constants, as defined in Eqs.~\eqref{eq:chi0} and \eqref{eq:kappa0}, respectively. $*$ denotes complex conjugation. }
    \label{tab2}
\end{table}

To obtain this first nonvanishing term of $\sigma^{\alpha}_{ij}(\bm{q})$,
\begin{align} \label{eq:sigmaDC_2}
    \sigma^{\alpha}_{ij}(\bm{q}) \simeq & \; \frac{i}{\pi}\frac{g^2e\hbar^6}{m^2M_{ph}}\int\frac{d^3p}{(2\pi)^3}\int\frac{d^3k}{(2\pi)^3} \frac{n_B(\omega_{\bm{k}})+n_F(\omega_{\bm{k}})}{2\omega_{\bm{k}}}  \nonumber \\
    &\times p_i^2\big[-\bm{q}\cdot(\hat{\bm{e}}_i\times\hat{\bm{e}}_\alpha)(\bm{q}\cdot\hat{\bm{e}}_j) + \bm{q}\cdot(\hat{\bm{e}}_j\times\hat{\bm{e}}_\alpha)(\bm{q}\cdot\hat{\bm{e}}_i)\big](1-\hat{k}_\alpha^2) |G_{\textrm{ret}}(0, \bm{p})|^4 \nonumber \\
     & + \frac{i}{\pi} \frac{g^2e\hbar^6}{m^2M_{ph}} \int\frac{d^3p}{(2\pi)^3}\int\frac{d^3k}{(2\pi)^3} \frac{n_B(\omega_{\bm{k}})+n_F(\omega_{\bm{k}})}{2\omega_{\bm{k}}} \nonumber \\
    & \times p_i p_j \big[\bm{q} \cdot(\bm{p} \times \hat{\bm{e}}_\alpha)\big] (1-\hat{k}_\alpha^2)\frac{4\hbar^2}{m}(\bm{p} \cdot \bm{q})|G_{\textrm{ret}}(0, \bm{p})|^4 G_{\textrm{ret}}(0, \bm{p}) + \mathcal{O}(q^4),
\end{align}
we need to consider how the electron propagator should be expanded in the powers of $\bm{q}$ and how the dependence on $\bm{k}$ should be approximated; given that we are interested in the long wavelength limit, our expansion will assume $|\bm{q}|l\ll1$, where $l=v_F\tau$ is the mean-free-path (see Supplementary Note 6 for further details).  In this equation, the first term is derived from Eq.~\eqref{eq:sigmaDC_1} by substituting $\bm{q}=0$ into the electron propagators and collecting nonvanishing even terms in $\bm{q}$. This term contributes to the imaginary part of the spin conductivity as the integrand is purely real.  But there is also the second term arising from expanding the electron propagators in $\bm{q}$. This expansion leads to the additional factor $\frac{4\hbar^2\bm{p}\cdot\bm{q}}{m}G_{\textrm{ret}}(0, \bm{p})$. Conventionally, this term is disregarded because it is of higher order in $\bm{q}$. However, in our case, both this term and the first term are quadratic in $\bm{q}$, and hence, both must be retained. The second term contributes to the real part of the spin conductivity as the additional propagator $G_{\textrm{ret}}(0, \bm{p})$ in the integrand introduces a purely imaginary factor $\big(-i\frac{E_F\tau}{\hbar}\big)$. It is noteworthy that this real part predominates in typical metals as $\frac{E_F\tau}{\hbar}\gg 1$ is required for having well-defined quasiparticles; conversely, close to the ferroelectric quantum criticality, the predominance of the real part cannot be taken for granted. It needs to be noted here that, in deriving Eq.~\eqref{eq:sigmaDC_2}, the phonon momentum $\bm{k}$ is set to be zero except for $[n_B(\omega_{\bm{k}})+n_F(\omega_{\bm{k}})]/\omega_{\bm{k}}$ that effectively constrains $|\bm{k}|$ at small values $|\bm{k}| \ll |\bm{p}| \sim k_F$ in the $\bm{k}$-integral \cite{mahan}. Further details can be found in Methods.

The lowest-order $\bm{q}$-dependence and the temperature dependence of $\sigma^\alpha_{ij}(\bm{q})$ can be obtained analytically by performing the Eq.~\eqref{eq:sigmaDC_2} integration, details of which can be found in the Methods. The result can be written in the form
\begin{align} \label{eq:sigmaDC_3}
    \sigma^{\alpha}_{ij}(\bm{q}) = \chi_0\big(\nu_{ij}^\alpha(\bm{q}) +\nu_{ji}^\alpha(\bm{q})\big) + i\kappa_0 \big(-\nu_{ij}^\alpha(\bm{q}) +\nu_{ji}^\alpha(\bm{q})\big),
\end{align}
where $\nu_{ij}^\alpha(\bm{q})$ is a quadratic function of $\bm{q}$, defined as:
\begin{align}
    \nu_{ij}^\alpha(\bm{q})=\bm{q}\cdot(\hat{\bm{e}}_i\times\hat{\bm{e}}_\alpha)(\bm{q}\cdot\hat{\bm{e}}_j).
\end{align}
$\chi_{0}$ and $\kappa_{0}$ are $\bm{q}$-independent constants, which are defined as
\begin{align}
    \chi_0 & \equiv \bigg(\frac{ne\hbar\tau}{m}\bigg)\bigg(\frac{g^2}{\hbar M_{ph}c^3}\bigg)\bigg(\frac{k_BT\tau}{\hbar}\bigg)^2\bigg(\frac{E_F\tau}{\hbar}\bigg)\frac{8}{5\pi^2}I\bigg(\frac{E_g}{k_BT}\bigg), \label{eq:chi0} \\
    \kappa_0 & \equiv \bigg(\frac{ne\hbar\tau}{m}\bigg)\bigg(\frac{g^2}{\hbar M_{ph}c^3}\bigg)\bigg(\frac{k_BT\tau}{\hbar}\bigg)^2\frac{2}{3\pi^2}I\bigg(\frac{E_g}{k_BT}\bigg); \label{eq:kappa0}
\end{align}
which confirms $\chi_0/\kappa_0 = 4E_F \tau/15\hbar \gg 1$ when the quasiparticles are well-defined. In this expression, $n = \frac{k_F^3}{3\pi^2}$ and $E_F = \frac{\hbar^2 k_F^2}{2m}$ denote the electron density and the Fermi energy, respectively. $E_g$ denotes the phonon energy gap. $I\big(\frac{E_g}{k_BT}\big)$ denotes an integral function representing the integral over $|\bm{k}|$, which is given by: 
\begin{align}
    I(y) = \int^\infty_0dx\frac{x^2}{2\sqrt{x^2+y^2}\sinh{\big[\sqrt{x^2+y^2}\; \big]}} \approx \frac{\pi^2}{8} \exp\big(-ay^b\big), \label{eq:I_1}
\end{align}
where $a \approx 0.75$ and $b \approx 1.05$.

\subsection{Quadrupolar symmetry}
\label{sec:quad}

\begin{figure}[tb!]
    \centering   
    \includegraphics[width=.5\textwidth]{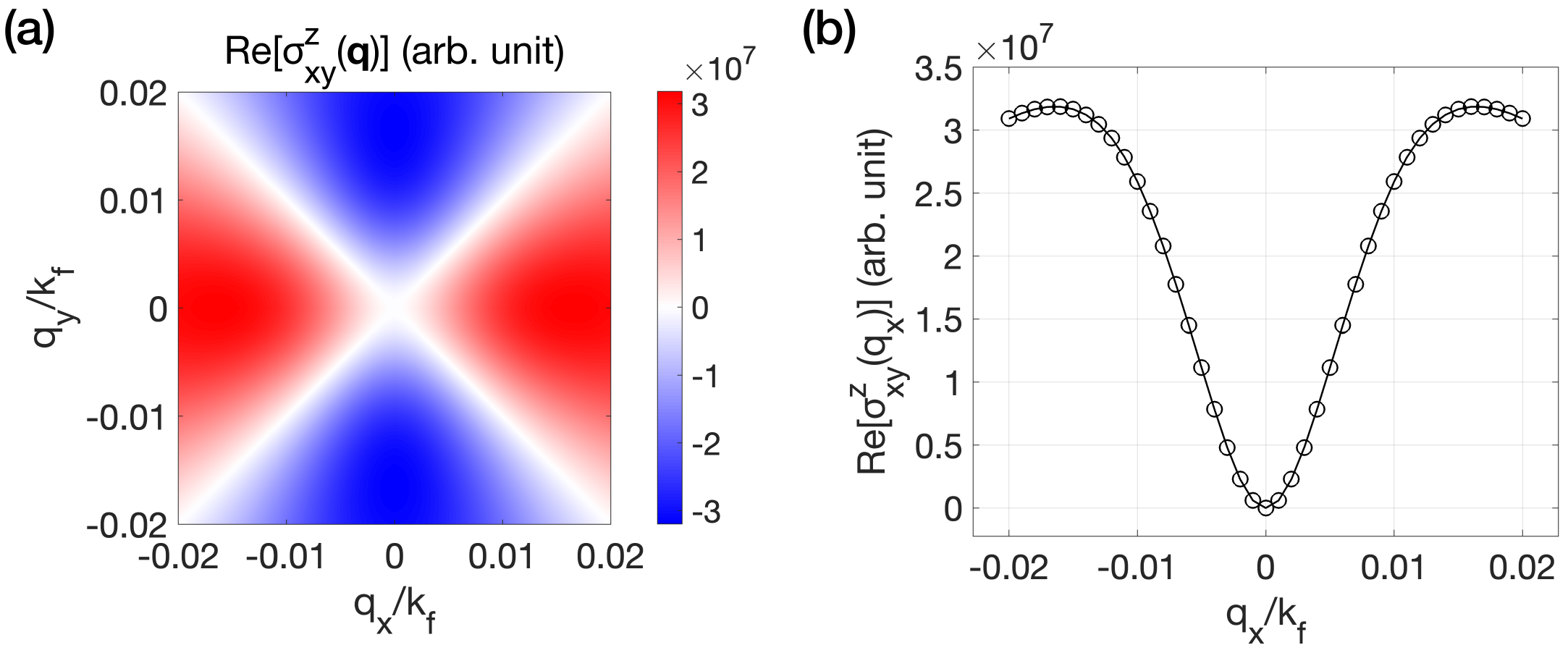}
    \caption[Quadrupole symmetry]{ \textbf{Quadrupole symmetry of the spin conductivity. } \textbf{a} Real part of the spin conductivity ($\textrm{Re}[\sigma_{xy}^z(\bm{q})]$) as a function of $\bm{q}=(q_x,q_y,0)$ (in units of $q_x/k_F$ and $q_y/k_F$) obtained from a numerical integration using the formula in Eq.~\eqref{eq:sigmaDC_1}.  \textbf{b} $\textrm{Re}[\sigma_{xy}^z(\bm{q})]$ along the line $q_y=0$, showcasing the quadratic dependence on $q_x^2$. \textbf{a} The color scale indicates the magnitude of the spin conductivity divided by the overall factor $\frac{g^2e\hbar^6}{m^2M_{ph}}$. \textbf{b} The parameter values used are as follows: $E_F=3\textrm{eV}$, $v_F=10^6\textrm{m/s}$, $\tau=10^{-14}\textrm{s}$, $c=10^4\textrm{m/s}$, $E_g=1\textrm{meV}$, and $T=1\textrm{K}$.}
    \label{fig4}
\end{figure}

The nonvanishing components of rank-3 tensor $\sigma_{ij}^\alpha(\bm{q})$ exhibit the unconventional quadrupolar symmetry in $\bm{q}$. In principle, $\sigma_{ij}^\alpha(\bm{q})$ could have twenty-seven distinct components, among various combinations of $(i, j, \alpha)$ where $i, j, \alpha = \{x, y, z\}$. However, three longitudinal components with $i = j = \alpha$ vanish, remaining only six components with $i = j \neq \alpha$. Moreover, we can see from Eq.~\eqref{eq:sigmaDC_3} that each pair of transverse components with $i \neq j$ is related by a symmetric relation in the permutation $i \leftrightarrow j$:
\begin{align} \label{eq:Re_sigma_sym}
    \sigma_{ij}^\alpha(\bm{q})=[\sigma_{ji}^\alpha(\bm{q})]^*,
\end{align}
where $^*$ denotes a complex conjugation. The symmetry of Eq.~\eqref{eq:Re_sigma_sym} also reduces the number of independent components to nine for the transverse components. The six longitudinal and nine transverse components are summarized in Table~\ref{tab1} and Table~ \ref{tab2}, respectively. These response functions collectively characterize the spin conductivity in the quantum paraelectric metal. Among these nonvanishing components, of particular interest are $\sigma_{xy}^z(\bm{q})$ and $\sigma_{yx}^z(\bm{q})$. These transverse spin conductivity components are explicitly given, in accordance with Eq.~\eqref{eq:Re_sigma_sym} as:
\begin{align}
\label{eq:Re_sigma_sym2}
    \sigma_{xy}^z(\bm{q}) & = \chi_0 (q_x^2 - q_y^2) + i\kappa_0 (q_x^2 + q_y^2), \\
    \label{eq:Re_sigma_sym3}
    \sigma_{yx}^z(\bm{q}) & = \chi_0 (q_x^2 - q_y^2) - i\kappa_0 (q_x^2 + q_y^2).
\end{align}
Note here the quadrupolar dependence on $\bm{q}$ characterized by $q_x^2-q_y^2$ for the real parts of these components. Our numerical computation on Eq.~\eqref{eq:sigmaDC_2} confirms this quadrupolar behavior within a low-$|\bm{q}|$ regime (Fig.~\ref{fig4}a).

The quadrupolar symmetry can be understood as a long-wavelength property dictated by the symmetry of the system. The absence of any odd power of $\bm{q}$ can be attributed to the inversion symmetry. Combined with the aforementioned absence of any spin conductivity constant in $\bm{q}$, this means that the spin conductivity in the long-wavelength limit $\bm{q} \to 0$ should be quadratic in $\bm{q}$. In addition, the rotational invariance of our model dictates that its response function, $\sigma_{ij}^{\alpha}(\bm{q})$ should also exhibit the rotational invariance around ${\bf \hat{e}}_\alpha$. As an illustration of how our quadrupolar symmetry satisfies the rotational invariance, let us consider the two components shown in Eqs.~\eqref{eq:Re_sigma_sym2} and \eqref{eq:Re_sigma_sym3}. The invariance under the $\pi/2$ rotation around the $z$-axis requires $\sigma_{xy}^{z}(q_x, q_y, q_z)=-\sigma_{yx}^{z}(-q_y, q_x, q_z)$. The quadrupolar symmetry of the real part proportional to $q_x^2-q_y^2$ generates the additional sign change under the $\pi/2$ rotation rotation of the wave vector $\bm{q}$ ($q_x\rightarrow-q_y$ and $q_y\rightarrow q_x$). This compensates for the sign between the two sides, ensuring the transformation rule is satisfied. 

\begin{figure}[tb!]
    \centering   
    \includegraphics[width=.5\textwidth]{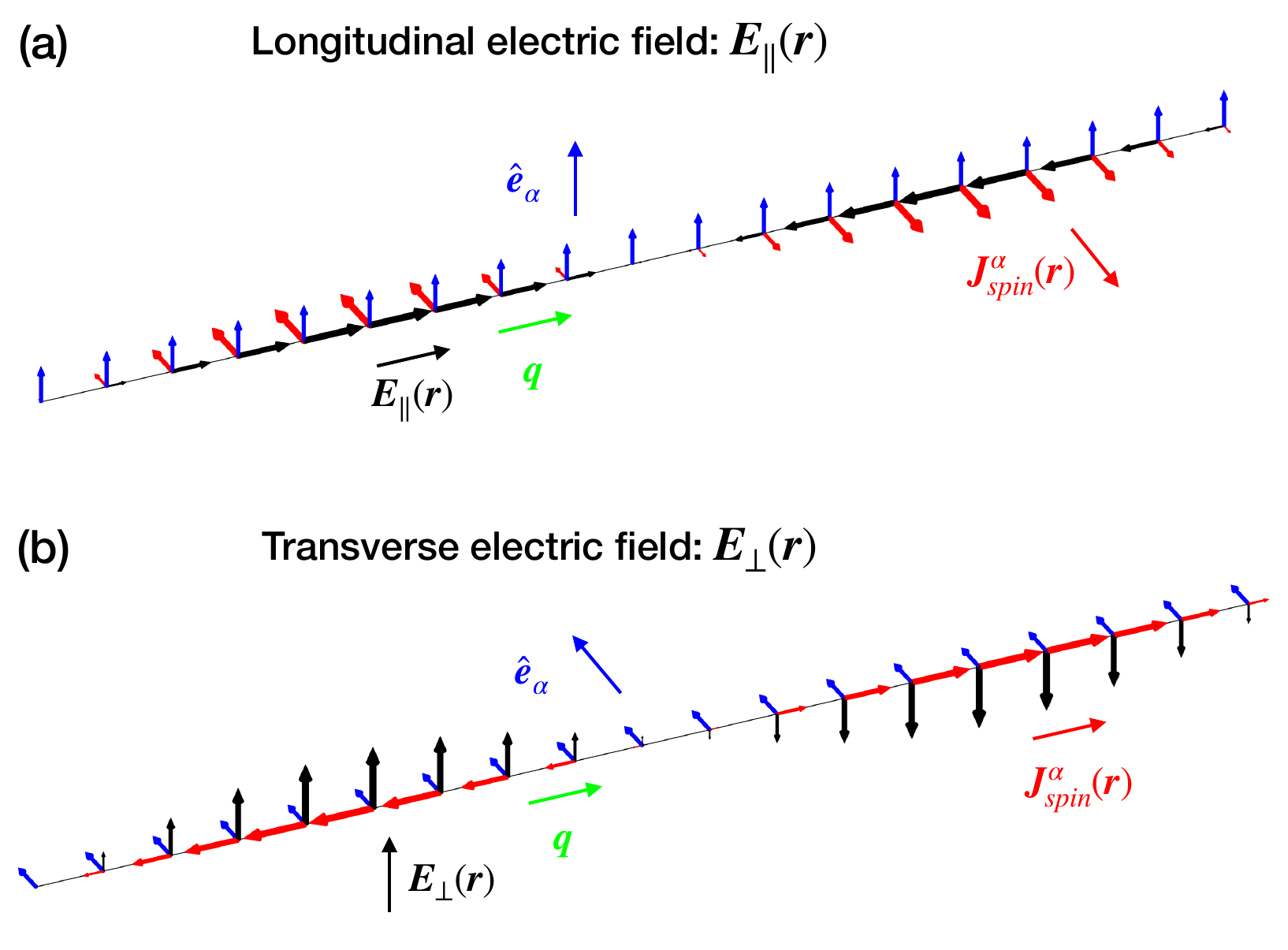}
    \caption{\textbf{Schematic illustration of spin currents generated by external electric fields.} \textbf{a} The longitudinal electric field ($\bm{E}_\parallel(\bm{r})$) and its wave vector ($\bm{q}$) align in the same $x$ direction. A sinusoidal form $\bm{E}_\parallel(\bm{r})=E_0\hat{x}\sin{(\bm{q}\cdot\bm{r})}$ is assumed where translation symmetry is assumed in the other directions. The spin current ($\bm{J}^\alpha_\textrm{spin}(\bm{r})$) and the spin quantization axis ($\hat{\bm{e}}_\alpha$) point in the $y$ and $z$ directions, respectively. \textbf{b} The transverse electric field ($\bm{E}_\perp(\bm{r})$) and its wave vector ($\bm{q}$) point in the different $z$ and $x$ directions. A sinusoidal form $\bm{E}_\perp(\bm{r})=E_0\hat{z}\sin{(\bm{q}\cdot\bm{r})}$ is assumed. The spin current ($\bm{J}^\alpha_\textrm{spin}(\bm{r})$) and spin quantization axis ($\hat{\bm{e}}_\alpha$) point in the $y$ and $z$ directions, respectively. In each plot, the thin black line guides the real-space modulation of $\bm{E}_\parallel(\bm{r})$ and $\bm{E}_\perp(\bm{r})$ along the $x$ direction. The sizes of the black and red arrows indicate the magnitudes of the electric field and the spin current, respectively. \textbf{a} The longitudinal electric field and its wave vector are indicated by the black and green arrows, respectively. The spin current and the spin quantization axis are shown, respectively, by the red and blue arrows. \textbf{b} The transverse electric field and its wave vector are denoted by the black and green arrows, respectively. The spin current and the spin quantization axis are denoted by the red and blue arrows, respectively.}
	\label{fig5}
\end{figure}

The phonon-mediated spin conductivity can also be written as the sum of very distinct responses to the longitudinal and the transverse electric field, respectively, allowing us to clarify its quadrupolar symmetry further. This can be derived by using $\sigma_{ij}^\alpha(\bm{q})$ in Eq.~\eqref{eq:sigmaDC_3} to calculate spin currents generated by external electric fields, which are expressed as:
\begin{align} \label{eq:spin_current_1}
    \bm{J}_{\textrm{spin}}^\alpha(\bm{q}) = \sum_{i,j}\hat{\bm{e}}_i\sigma_{ij}^\alpha(\bm{q}) E_{j}(\bm{q}).
\end{align}
Here, $E_{j}(\bm{q})$ denotes an electric field applied in the $j$ direction. Substituting Eq.~\eqref{eq:sigmaDC_3} into Eq.~\eqref{eq:spin_current_1}, we obtain the first of our main results, an explicit expression of $\bm{J}_{\textrm{spin}}^\alpha(\bm{q})$:
\begin{align} \label{eq:spin_current_2} 
    \bm{J}_{\textrm{spin}}^\alpha(\bm{q}) = (\chi_0 - i\kappa_0) (\hat{\bm{e}}_\alpha\times\bm{q})[\bm{q} \cdot \bm{E}_\parallel(\bm{q})] + (\chi_0 + i\kappa_0) \bm{q}[\hat{\bm{e}}_\alpha \cdot(\bm{q} \times \bm{E}_\perp(\bm{q}))].
\end{align}
Here, $\bm{E}_\perp (\bm{q})$ and $\bm{E}_\parallel (\bm{q})$ represent longitudinal and transverse components of $\bm{E}(\bm{q})$ with respect to $\bm{q}$, respectively, which are defined as
\begin{align}
    \bm{E}_\parallel (\bm{q}) & = \frac{\bm{q}(\bm{q} \cdot \bm{E} (\bm{q}))}{|\bm{q}|^2}, \label{eq:E_longi} \\
    \bm{E}_\perp (\bm{q}) & = \bm{E}(\bm{q}) - \frac{\bm{q}(\bm{q} \cdot \bm{E} (\bm{q}))}{|\bm{q}|^2}.
    \label{eq:E_trans} 
\end{align}
We note that the spin current of Eq.~\eqref{eq:spin_current_2} will be non-uniform in general, {\it e.g.} when $\kappa_0/\chi_0$ can be ignored, it can be written in the real space approximately as
\begin{equation*}
\bm{J}_{\textrm{spin}}^\alpha = -\chi_0\bigg[(\hat{\bm{e}}_\alpha\times\bm{\nabla})(\bm{\nabla}\cdot\bm{E})+\bm{\nabla}\{\hat{\bm{e}}_\alpha\cdot(\bm{\nabla}\times\bm{E})\}\bigg].
\end{equation*}
Eq.~\eqref{eq:spin_current_2} tells us that $\bm{E}_\parallel$ generates the spin Hall current perpendicular to the spin polarization direction $\hat{\bm{e}}_\alpha$ whereas $\bm{E}_\perp$ generates the $\bm{q}$-parallel spin Hall current whose magnitude is maximized for $\hat{\bm{e}}_\alpha \perp \bm{q}$. For an example of response when $\bm{E} = \bm{E}_\parallel$, we can consider  $\bm{q}=q\hat{\bm{x}}$, $\bm{E}(\bm{q})=E_0\hat{\bm{x}}$, and $\hat{\bm{e}}_\alpha=\hat{\bm{z}}$, for which the resulting spin Hall current flows in the $y$ direction (Fig.~\ref{fig5}a):
\begin{align} \label{eq:spin_current_longit}
    \bm{J}_{\textrm{spin}}^z(\bm{q}) = (\chi_0 - i\kappa_0) \hat{\bm{y}}q^2E_0.
\end{align}
In fact, this spin current is of the form analogous to the charge current in quantum Hall states induced by inhomogeneous electric field \cite{haldane2009hall, PhysRevLett.108.066805,
Avron1995, Read2009}; the spin analogue in the quantum spin Hall state \cite{Kimura2021} and the Rashba metal \cite{Zhang2022} has also been discussed. For an example of response when $\bm{E}=\bm{E}_\perp$, we can consider $\bm{q}=q\hat{\bm{x}}$, $\bm{E}(\bm{q})=E_0\hat{\bm{z}}$, and $\hat{\bm{e}}_\alpha=\hat{\bm{y}}$, for which the resulting spin current flows in the $x$ direction (Fig.~\ref{fig5}b):
\begin{align} \label{eq:spin_current_transv}
    \bm{J}_{\textrm{spin}}^y(\bm{q}) = (\chi_0+i\kappa_0)(-\hat{\bm{x}})q^2E_0.
\end{align}
This $\bm{q}$-parallel Hall spin current induced by $\bm{E}_\perp$ has, to the best of our knowledge, no analogue reported in the transport of either the Rashba metal or the quantum Hall state and can be considered as the most distinct feature of the phonon-mediated spin conductivity.

The above separation of the two components of the spin conductivity is physically relevant as a metal exhibits different responses to $\bm{E}_\parallel$ and $\bm{E}_\perp$. Briefly, $\bm{E}_\parallel$ is screened below the plasma frequency, whereas, as dictated by the Faraday effect, $\bm{E}_\perp$ is necessarily dynamic; note that the coupling of electric field to the TO phonon, omitted in our analysis, can be adequately treated with the static dielectric constant for frequency far below $E_g/\hbar$. Indeed, the response to $\bm{E}_\perp$ can be interpreted as the generation of spin current in response to an electromagnetic wave if we additionally take into account the frequency dependence of the spin conductivity. For instance, we have derived in 
Supplementary Notes 4 and 7 the approximate relation, valid for the low frequency $\omega \tau \ll 1$, between the static and dynamic conductivity in the first order in $\omega$:
\begin{align}
    \sigma_{ij}^\alpha(\omega, \bm{q}) = \bigg(1 + \frac{3}{2}i\omega\tau\bigg) \sigma_{ij}^\alpha(\bm{q}).
\end{align}
Physically, large differences between $|\bm{E}_\parallel|$ and $|\bm{E}_\perp|$ is required for obtaining a nearly pure spin Hall current; $|\bm{E}_\parallel(\bm{q})|\neq0$ and $|\bm{E}_\perp(\bm{q})|\neq0$, on the other hands, will give us a nonzero longitudinal spin current. For instance, with $\bm{E}(\bm{q}=q\hat{\bm{x}}) = E_0\frac{\hat{\bm{x}} + \hat{\bm{z}}}{\sqrt{2}}$ and $\hat{\bm{e}}_\alpha=\hat{\bm{y}}$, the spin Hall and longitudinal currents are given by
\begin{align}
    \bm{J}_\textrm{sH}^y(\bm{q}=q\hat{\bm{x}}) & = -i\kappa_0 \frac{\hat{\bm{x}} - \hat{\bm{z}}}{\sqrt{2}} q^2 E_0, \\
    \bm{J}_\textrm{lon}^y(\bm{q}=q\hat{\bm{x}}) & = -\chi_0 \frac{\hat{\bm{x}} + \hat{\bm{z}}}{\sqrt{2}} q^2 E_0.
\end{align}
Likewise, for any of the longitudinal spin conductivity components of Table~\ref{tab1} to be nonzero, both $|\bm{E}_\parallel|\neq0$ and $|\bm{E}_\perp|\neq0$ need to hold.

One relevant parameter for assessing the magnitude of the spin conductivity in Eq.~\eqref{eq:sigmaDC_3} is the `spin Hall angle' defined as $\theta_\textrm{sH} \equiv \frac{2e}{\hbar} \frac{|\bm{J}_\textrm{spin}^{\alpha}(\bm{q})|}{|\bm{J}_\textrm{e}(\bm{q})|}$, where $\bm{J}_\textrm{e}(\bm{q}) = \sigma_\textrm{e}(\bm{q}) \bm{E}(\bm{q}) $ is the electric current induced by the same external electric field $\bm{E}(\bm{q})$. Using the Drude conductivity $\sigma_\textrm{e}(\bm{q}) = \frac{ne^2\tau}{m}$ with assuming $\sigma_\textrm{e}(\bm{q})$ independent of $\bm{q}$, we obtain $\theta_\textrm{sH} = \frac{2e}{\hbar} \frac{\sqrt{\chi_0^2+\kappa_0^2}}{\sigma_{e}} q^{2} F_{\alpha}(\hat{\bm{q}}, \hat{\bm{E}})$ or
\begin{align}
    \theta_\textrm{sH} = \bigg(\frac{g^{2}q^{2}}{\hbar M_{ph}c^{3}}\bigg)\bigg(\frac{k_BT\tau}{\hbar}\bigg)^2I\bigg(\frac{E_g}{k_BT}\bigg) \sqrt{\bigg(\frac{8}{5\pi^2}\bigg)^2\bigg(\frac{E_F\tau}{\hbar}\bigg)^2 + \bigg(\frac{2}{3\pi^2}\bigg)^2} F_{\alpha}(\hat{\bm{q}}, \hat{\bm{E}}),
\end{align}
where the directional factor $F_{\alpha}(\hat{\bm{q}}, \hat{\bm{E}}) = \sqrt{(\hat{\bm{q}}\cdot\hat{\bm{E}})^2 + ((\hat{\bm{q}}\times\hat{\bm{e}}_\alpha)\cdot\hat{\bm{E}})^2}$ is order of unity. To provide a rough numerical estimate, we consider SrTiO\textsubscript{3} as an example despite its deviations from our model, including its multi-orbital electronic structure. Employing the following parameter values: $g\approx30 \textrm{ meV}a^{3/2}$, with $a \approx 5.5$ \AA{} being the lattice constant of tetragonal SrTiO\textsubscript{3} \cite{PhysRevResearch.5.023177}, $E_g \approx 2$ meV and $E_F \approx 10$ meV for $n=4\times10^{19}\textrm{cm}^{-3}$ \cite{yoon2021lowdensity}, $\tau=10^{-11}$ s for $T=10$ K \cite{Mikheev2016}, and assuming $M_{ph} \approx 4\times10^{-26}$ kg and $c=10^3$ m/s for the TO$_1$ \cite{PhysRevResearch.5.023177}, we obtain $\theta_\textrm{sH} \approx 8.36 \times 10^{-16} q^{2} \textrm{m}^{2}$. Specifically, for $q=(3\times10^{-7}\textrm{m})^{-1}$, this yields $\theta_\textrm{sH} \approx 0.0093$, which is comparable to that of Pt at the same temperature. It's notable that the relatively high value of $k_{B}T\tau/\hbar \approx 13$ and its quadratic dependence are primarily responsible for the substantial $\theta_\textrm{sH}$.

\subsection{Thermal activation behavior}
\label{sec:thermal}

\begin{figure}[tb!]
    \centering
    \includegraphics[width=.5\textwidth]{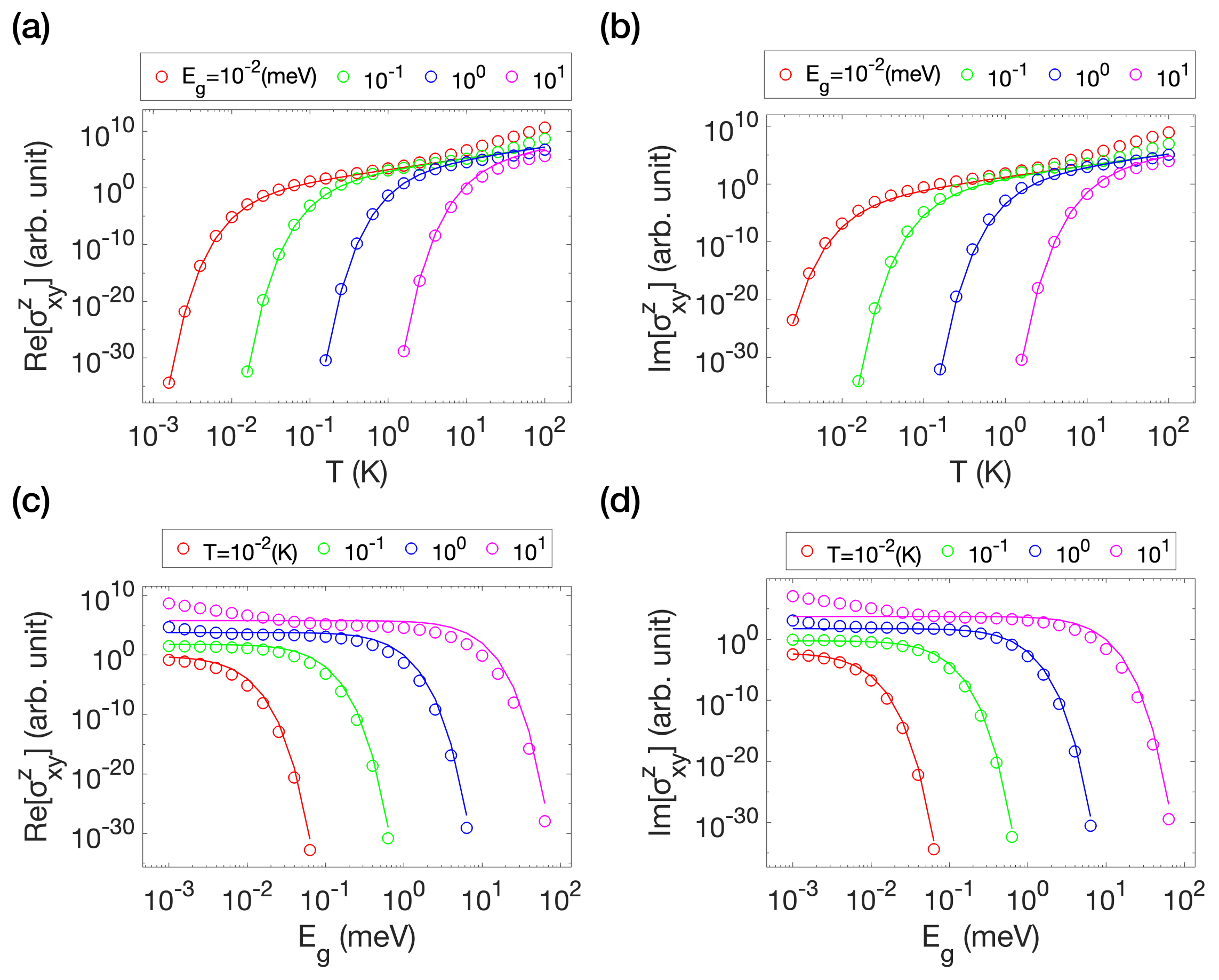}
    \caption{\textbf{Scaling laws in spin conductivity.} \textbf{a} Real part of the spin conductivity ($\textrm{Re}[\sigma_{xy}^z(\bm{q})]$) as a function of temperature ($T$). \textbf{b} Imaginary part of the spin conductivity ($\textrm{Im}[\sigma_{xy}^z(\bm{q})]$) as a function of $T$. Different curves represent distinct values of the energy gap ($E_g$). \textbf{c} Real part of the spin conductivity ($\textrm{Re}[\sigma_{xy}^z(\bm{q})]$) as a function of $E_g$. \textbf{d} Imaginary part of the spin conductivity ($\textrm{Im}[\sigma_{xy}^z(\bm{q})]$) as a function of $E_g$. Different curves represent distinct values of $T$. In each plot, the dashed line represents the theoretical curve for the spin conductivity based on the analytical formula in Eq.~\eqref{eq:sigmaDC_1}. Additionally, the magnitude of the spin conductivity is normalized by the overall factor $\frac{g^2e\hbar^6}{m^2M_{ph}}$. The parameter values used are as follows: $E_F=3\textrm{eV}$, $v_F=10^6\textrm{m/s}$, $\tau=10^{-14}\textrm{s}$, $c=10^4\textrm{m/s}$, and $\bm{q}=(0.001k_f,0,0)$.}
    \label{fig6}
\end{figure}

The second of our main results is the scaling laws with respect to temperature governing the phonon-mediated spin conductivity, as represented by:
\begin{align} \label{eq:SHC_scaling}
    \sigma_{ij}^\alpha(\bm{q}) \propto |\bm{q}|^2 \bigg(\frac{k_BT\tau}{\hbar}\bigg)^2 \exp\Bigg[-a\bigg(\frac{E_g}{k_BT}\bigg)^b\Bigg],
\end{align}
where $a \approx 0.75$ and $b \approx 1.05$, as obtained in Eq.~\eqref{eq:I_1}. Importantly, $\sigma_{ij}^\alpha(\bm{q})$ becomes zero at absolute zero temperature ($T=0$). Conversely, the spin conductivity undergoes enhancement as the temperature increases. This behavior stems from the mediation of thermally excited TO phonons. This thermal activation behavior stands in stark contrast to the previous intrinsic spin Hall conductivity, which maintains a nonzero value at the zero temperature \cite{PhysRevLett.92.126603}. It also represents a significant departure from the impact of acoustic phonons, which lead to the degradation of electrical conductivity \cite{Ashcroft76}. It is worth noting that the quadratic temperature dependence in the scaling formula originates from the phase volume factor $k_\textrm{cutoff}^3 = \big(\frac{k_BT}{\hbar c}\big)^3$ divided by an additional factor of $k_BT$ stemming from the phonon energy $\omega_{\bm{k}}$. Roughly speaking, this factor accounts for the density of thermally excited TO phonons that give rise to the generation of the PMSP. Additionally, the presence of four electron propagators results in a cubic dependence on the electron's scattering time $\tau$, another crucial departure from the usual linear dependence observed in conductivity.

When $E_g \ll k_BT$, as indicated by the red region in Fig.~\ref{fig1}, TO phonons exhibit gapless behavior, and thermal effects dominate. In this gapless-phonon regime, the exponential factor can be disregarded, simplifying the scaling formula in Eq.~\eqref{eq:SHC_scaling} as follows:
\begin{align} \label{eq:SHC_scaling_quad}
    \sigma_{ij}^\alpha(\bm{q}) \propto |\bm{q}|^2 \left(\frac{k_BT\tau}{\hbar}\right)^2, \quad (E_g \ll k_B T).
\end{align}
Our numerical results confirm this power-law relationship, as depicted in Fig.~\ref{fig6}a–b. This quadratic temperature scaling is a distinctive feature of the phonon-mediated spin conductivity, specifically in the gapless-phonon regime.

However, it needs to be recognized that as temperatures decrease, the system eventually enters the quantum critical regime $(E_g \ll k_BT \lesssim k_BT_*)$, as indicated by the blue region in Fig.~\ref{fig1}. 
In this regime, our results may not be applicable due to the significant impact of phonon self-energy corrections, invalidating the usage of the free phonon propagator, which forms the basis of our calculations. Our analysis of phonon self-energy corrections provides an estimate of $T_* \sim \frac{E_F}{k_B} \sqrt{\frac{3\pi}{2}\frac{g^2n}{E_F^2}\frac{mc}{M_{ph}v_F}}$ (see 
Supplementary 
Note 9 for the derivation of $T_*$). Consequently, the quadratic scaling law for phonon-mediated spin conductivity remains valid within a limited temperature range $T > T_*$.

Conversely, when $E_g \gg k_BT$, the exponential factor dominates, and the power-law component can be neglected. In such cases, the scaling formula in Eq.~\eqref{eq:SHC_scaling} can be expressed simply as:
\begin{align}
    \sigma_{ij}^\alpha(\bm{q}) \propto |\bm{q}|^2 \exp\left[-a\left(\frac{E_g}{k_BT}\right)^b\right], \quad (E_g \gg k_B T).
\end{align}
Our numerical results confirm this exponential behavior in the gapped-phonon regime, as shown in Fig.~\ref{fig6}c, d. This thermal activation behavior emphasizes that the spin Hall effect is facilitated by thermally excited TO phonons, and the presence of TO phonons becomes increasingly favorable as their energy gap diminishes, enhancing the spin conductivity. We suggest that this scaling law is applicable in the “classical” paraelectric phase near the quantum critical point where $E_g \gg k_BT$, as depicted by the yellow region in Fig.~\ref{fig1}.

In principle, our calculations can be extended to the quantum critical regime by accounting for self-energy effects. In this scenario, the detailed characteristics of the spin conductivity, such as the quadratic scaling law, may undergo potential modifications. Nevertheless, the presence of the spin conductivity remains robust since the mediation of the spin conductivity by TO phonons, as described by the PMSP in Eq.~\eqref{eq:gamma_tilde_1}, sorely relies on Rashba-type spin-orbit coupling, which is a fundamental aspect of the system, irrespective of the specific phonon or electron propagator. Consequently, we argue that the spin conductivity persists even in the quantum critical regime, though it may not necessarily adhere to the specific form, as presented in Eq.~\eqref{eq:SHC_scaling}, and could manifest with a different temperature dependence characterized by a different exponent and numerical coefficients.

\subsection{Anisotropy effects} \label{sec:anisotropy}

Many metals near the ferroelectric phase transition exhibit quasi-two-dimensional behavior or belong to the category of two-dimensional (2D) materials \cite{Zhou_2020, PhysRevMaterials.7.010301, Bhowal2023}; the possibility in van der Waals materials is attracting widespread interest in recent years \cite{Wang2023} with the discovery of WTe$_2$ metallic ferroelectricity \cite{Fei2018, Yang2018} and the prediction of strain-tunable ferroelectricity in $\beta$-GeSe \cite{Guan2018} being just two of many examples. The crystalline anisotropy in such systems significantly impacts the energy dispersion of TO phonons, leading to either an easy-plane or easy-axis behavior. Additionally, the anisotropic character can be experimentally manipulated using external stimuli, such as strain \cite{Haeni2004}. Furthermore, the electron-phonon coupling effect on the electrons' dynamics is quite different in 2D anisotropic systems compared to 3D isotropic systems \cite{PhysRevB.107.165110}. Hence, it's crucial to consider anisotropy when applying our theoretical model to the candidate materials for the quantum paraelectric metal.

Under these considerations, we extend our previous three-dimensional (3D) isotropic results, as presented in Eqs. \eqref{eq:sigmaDC_1} and \eqref{eq:sigmaDC_3}, to encompass anisotropic 2D scenarios, including easy-plane and easy-axis cases. The primary modification arises in the factor $\tilde{\gamma}^\alpha(\bm{p},\bm{k};\bm{q})$ in Eq.~\eqref{eq:gamma_tilde_1}, which is adjusted as follows:
\begin{align}
    \label{eq:gamma_EP}
    \tilde{\gamma}^{\alpha,\textrm{EP}}(\bm{p},\bm{k};\bm{q}) & = (1-\delta_{\alpha z}) \tilde{\gamma}^\alpha(\bm{p},\bm{k};\bm{q}), \\
    \tilde{\gamma}^{\alpha,\textrm{EA}}(\bm{p},\bm,k;\bm{q}) & =  \delta_{\alpha z} \tilde{\gamma}^\alpha(\bm{p},\bm{k};\bm{q}).
    \label{eq:gamma_EA}
\end{align}
Here, the superscripts “EP” and “EA” denote “easy-plane” and “easy-axis,” respectively. In comparison to the isotropic case, $\tilde{\gamma}^{\alpha}(\bm{p},\bm{k};\bm{q})$ now exhibits directional preferences denoted by projection factors such as $(1-\delta_{\alpha z})$ and $\delta_{\alpha z}$. These projection factors constrain the direction of the spin polarization to the easy-plane $\alpha=\{x,y\}$ or to the easy-axis $\alpha=z$ in each case. This spin polarization constraint clearly differentiates the spin conductivity in anisotropic systems from that of isotropic systems.

To precisely quantify the deviations from the isotropic scenario, we calculate the phonon-mediated for both easy-plane and easy-axis cases, which are denoted as $\sigma^{\alpha,\textrm{EP}}_{ij}(\bm{q})$ and $\sigma^{\alpha,\textrm{EA}}_{ij}(\bm{q})$, respectively, and find the same quadrupolar symmetry with the different response for the transverse and the longitudinal field, albeit with the spin polarization constraint of Eqs.~\eqref{eq:gamma_EP} and \eqref{eq:gamma_EA}. The calculations closely mirror those of the isotropic case; for further details, refer to 
Supplementary 
Note 8. As a result, we obtain $\sigma^{\alpha,\textrm{EP}}_{ij}(\bm{q})$ and $\sigma^{\alpha,\textrm{EA}}_{ij}(\bm{q})$ up to quadratic order in $\bm{q}$ as:
\begin{align} \label{eq:sigma_aniso}
    \sigma^{\alpha, \textrm{EP}}_{ij}(\bm{q}) = & \; (1-\delta_{\alpha z})\Big[ \chi_0^\textrm{2D} \big(\nu_{ij}^\alpha(\bm{q}) +\nu_{ji}^\alpha(\bm{q})\big) + i\kappa_0^\textrm{2D} \big(-\nu_{ij}^\alpha(\bm{q}) +\nu_{ji}^\alpha(\bm{q})\big) \Big], \\
    \sigma^{\alpha, \textrm{EA}}_{ij}(\bm{q}) = & \; \delta_{\alpha z} \Big[ \chi_0^\textrm{2D} \big(\nu_{ij}^\alpha(\bm{q}) + \nu_{ji}^\alpha(\bm{q})\big) + i\kappa_0^\textrm{2D} \big(-\nu_{ij}^\alpha(\bm{q}) + \nu_{ji}^\alpha(\bm{q})\big) \Big].
\end{align}
In contrast to the isotropic case, the spin quantization direction $\alpha$ cannot be freely adjusted any longer; for the easy-plane case, $\alpha$ is constrained to the easy-plane directions, while for the easy-axis case, it aligns with the easy-axis direction as indicated by the factors $1-\delta_{\alpha z}$ and $\delta_{\alpha z}$, respectively. For the easy axis case, the spin current response to the longitudinal electric field is now exactly the same form as that of the Rashba metal \cite{Zhang2022}.

The qualitative change due to the 2D nature does occur in the temperature dependence, as can be seen from the coefficients of the spin conductivity,
\begin{align}
    \chi_0^\textrm{2D} & = \bigg(\frac{n_\textrm{2D}e\hbar\tau}{m}\bigg)\bigg(\frac{g^2\tau}{\hbar M_{ph}c^2}\bigg)\bigg(\frac{k_BT\tau}{\hbar}\bigg) \bigg(\frac{E_F\tau}{\hbar}\bigg)\frac{3}{2\pi}I_\textrm{2D}\bigg(\frac{E_g}{k_BT}\bigg), \label{eq:chi0_2D} \\
    \kappa_0^\textrm{2D} & = \bigg(\frac{n_\textrm{2D}e\hbar\tau}{m}\bigg)\bigg(\frac{g^2\tau}{\hbar M_{ph}c^2}\bigg)\bigg(\frac{k_BT\tau}{\hbar}\bigg)\frac{1}{2\pi}I_\textrm{2D}\bigg(\frac{E_g}{k_BT}\bigg), \label{eq:kappa0_2D}
\end{align}
where $n_\textrm{2D} = \frac{k_F^2}{2\pi}$ represents the electron density in 2D. It is $I_\textrm{2D}\big(\frac{E_g}{k_BT}\big)$, an integral function that encapsulates the integration over $|\bm{k}|$,
\begin{align}
    I_\textrm{2D}(y) = \int^\infty_0dx\frac{x}{2\sqrt{x^2+y^2}\sinh{\big[\sqrt{x^2+y^2}\; \big]}}=\frac{1}{2}\ln\bigg(\coth\frac{y}{2}\bigg), \label{eq:I_1_2D}
\end{align}
that gives rise to the most important qualitative change in the temperature and the phonon energy gap dependence. In the gapless-phonon regime, $E_g \ll k_BT$ or $y\ll1$, (the red region in Fig.~\ref{fig1}), $I_\textrm{2D}(y)$ can be well-approximated as $I_\textrm{2D}(y)\approx\frac{1}{2}\ln\big(\frac{2}{y}\big)$, simplifying the scaling formula in Eq.~\eqref{eq:sigma_aniso} as follows:
\begin{align} \label{eq:sigma_aniso_scal1}
    \sigma_{ij}^{\alpha,\textrm{EP(EA)}}(\bm{q}) \propto |\bm{q}|^2 \left(\frac{k_BT\tau}{\hbar}\right)\ln\bigg(\frac{2k_BT}{E_g}\bigg), \quad (E_g \ll k_B T).
\end{align}
Conversely, in the gapped-phonon regime, $E_g \gg k_BT$ or $y\gg1$, (the yellow region in Fig.~\ref{fig1}), $I_\textrm{2D}(y)$ can be approximated as $I_\textrm{2D}(y)\approx\exp(-y)$, simplifying the scaling formula in Eq.~\eqref{eq:sigma_aniso} as follows:
\begin{align} \label{eq:sigma_aniso_scal2}
    \sigma_{ij}^{\alpha,\textrm{EP(EA)}}(\bm{q}) \propto |\bm{q}|^2 \exp\left[-\left(\frac{E_g}{k_BT}\right)\right], \quad (E_g \gg k_B T).
\end{align}
In both easy-axis and easy-plane cases, the temperature dependence in Eqs.~(45) and (46) may undergo substantial modifications at the QCP where the effect of self-corrections becomes significant. Nevertheless, our results raise the possibility of stronger phonon-mediated spin conductivity in anisotropic 2D materials compared to isotropic 3D materials such as SrTiO$_3$.

\section{Discussion}

In this study, we have demonstrated that quantum paraelectric metals near ferroelectric quantum criticality can exhibit an unconventional phonon-mediated spin transport in response to an inhomogeneous electric field, an example of spin transport arising from interaction rather than band structure. Our rigorous calculations, employing the Kubo formula and a perturbative expansion in electron-phonon interaction, have unveiled that soft transverse optical phonons, with their intrinsic Rashba-type spin-orbit coupling to electrons, can serve as unconventional contributors to the spin conductivity in response to the inhomogeneous electric field. Furthermore, the resulting spin conductivity displays a couple of intriguing and unique characteristics. One is that it exhibits unconventional quadrupolar symmetry associated with the $\bm{q}$-vector, leading to a possible nonzero response, in contrast to the theoretical prediction for quantum Hall states and Rashba metals, even when $\bm{\nabla} \cdot \bm{E} = 0$. The resulting spin current is, therefore, non-uniform, and its observation may require a local spin probe such as the X-ray magnetic circular dichroism \cite{Stohr1999} or the spin-torque transfer ferromagnetic resonance \cite{Wang2018}; recent years have seen the successful application of both to the 2D van der Waals materials \cite{Khan2019, Ghosh2023}. The other is that it follows distinctive scaling laws in temperature and phonon energy gap; the conductivity increases as temperature rises or the energy gap diminishes, as it is mediated by thermally excited phonons. Consequently, the proposed spin transport may be best observed in the vicinity of a ferroelectric quantum criticality, where the phonon energy gap diminishes while thermal effects amplify, even if how much of Fermi liquid theory-based temperature dependence would hold in the quantum critical regime remains to be examined in future research. Therefore, the recent report of the ferroelectric quantum critical point in the n-type SrTiO$_3$ may provide one example \cite{annurev:/content/journals/10.1146/annurev-conmatphys-031218-013144, Rischau2022, Tomioka2022}; 2D van der Waals materials may provide other examples of ferroelectric quantum criticality \cite{Wang2023, Guan2018} and hence enhanced interaction-induced spin transport. Lastly, we note that the impurity effect on the spin conductivity is an interesting issue to be addressed, {\it e.g.} whether it would be analogous to that of the Rashba metal spin Hall conductivity \cite{PhysRevLett.93.226602, PhysRevB.72.165316}.


\section{Methods} \label{sec:methods}


\subsection{Expansion of spin conductivity to the wave vector of electric fields}

To obtain the quadratic term of $\sigma^{\alpha}_{ij}(\bm{q})$ presented in Eq.~\eqref{eq:sigmaDC_2}, we expanded the full expression of Eq.~\eqref{eq:sigmaDC_1} to the wave vector $\bm{q}$ of electric fields. Utilizing the expansion of $G_\textrm{ret}(0,\bm{k}+\bm{q})$ to $\bm{q}$: 
\begin{align}
    G_\textrm{ret}(0,\bm{k}+\bm{q}) & = \; G_\textrm{ret}(0,\bm{k}) \bigg( 1-G_\textrm{ret}(0,\bm{k})\frac{\hbar^2 \bm{k}\cdot \bm{q}}{2m} - G_\textrm{ret}(0,\bm{k})\frac{\hbar^2\bm{q}^2}{2m} \bigg)^{-1} \nonumber \\
    & \approx G_\textrm{ret}(0,\bm{k}) + \bigg(\frac{\hbar^2 \bm{k}\cdot \bm{q}}{2m}+\frac{\hbar^2\bm{q}^2}{2m}\bigg)G_\textrm{ret}(0,\bm{k})^2 + \bigg(\frac{\hbar^2 \bm{k} \cdot \bm{q}}{2m}\bigg)^2G_\textrm{ret}(0,\bm{k})^3 + \mathcal{O}(q^3),
\end{align}
and a similar expression for $G_\textrm{ret}(\omega_{\bm{k}},\bm{p}+\bm{k}+\bm{q})$, we expanded the product of the electron propagators in Eq.~\eqref{eq:sigmaDC_1} as:
\begin{align}
    &G_{\textrm{ret}}(0, \bm{p}+\bm{q}) G_{\textrm{adv}}(0, \bm{p}) G_{\textrm{ret}}(\omega_{\bm{k}},\bm{p}+\bm{k}+\bm{q}) G_{\textrm{adv}}(\omega_{\bm{k}},\bm{p}+\bm{k}) + (\omega_{\bm{k}} \longleftrightarrow -\omega_{\bm{k}})\nonumber\\ = & \; |G_{\textrm{ret}}(0, \bm{p})|^2|G_{\textrm{ret}}(\omega_{\bm{k}},\bm{p}+\bm{k})|^2 \bigg\{1+ \frac{\hbar^2\bm{p}\cdot\bm{q}}{m} G_{\textrm{ret}}(0, \bm{p}) + \frac{\hbar^2(\bm{p}+\bm{k})\cdot\bm{q}}{m} G_{\textrm{ret}}(\omega_{\bm{k}}, \bm{p}+\bm{k}) \nonumber \\ 
    & + \frac{\hbar^2\bm{q}^2}{2m} G_{\textrm{ret}}(0, \bm{p}) + \frac{\hbar^2\bm{q}^2}{2m} G_{\textrm{ret}}(\omega_{\bm{k}}, \bm{p}+\bm{k}) + \bigg(\frac{\hbar^2\bm{p}\cdot\bm{q}}{m} \bigg)^2 [G_{\textrm{ret}}(0, \bm{p})]^2 +\bigg(\frac{\hbar^2(\bm{p}+\bm{k})\cdot\bm{q}}{m} \bigg)^2 [G_{\textrm{ret}}(\omega_{\bm{k}}, \bm{p}+\bm{k})]^2 \nonumber \\
    & + \bigg(\frac{\hbar^2\bm{p}\cdot\bm{q}}{m} \bigg)\bigg(\frac{\hbar^2(\bm{p}+\bm{k})\cdot\bm{q}}{m} \bigg) G_{\textrm{ret}}(0, \bm{p})G_{\textrm{ret}}(\omega_{\bm{k}}, \bm{p}+\bm{k}) + \mathcal{O}(q^3) \bigg\} + (\omega_{\bm{k}} \longleftrightarrow -\omega_{\bm{k}}). \label{eq:G_exp}
\end{align}
Substituting Eq.~\eqref{eq:G_exp} into Eq.~\eqref{eq:sigmaDC_1} and gathering terms of quadratic order in $\bm{q}$, we obtain
\begin{align} \label{method:sigma1} 
    \sigma^{\alpha}_{ij}(\bm{q}) = & \; \frac{i}{\pi}\frac{g^2e\hbar^6}{m^2M_{ph}}\int\frac{d^3p}{(2\pi)^3}\int\frac{d^3k}{(2\pi)^3} \frac{n_B(\omega_{\bm{k}})+n_F(\omega_{\bm{k}})}{2\omega_{\bm{k}}} \Big\{\bm{q}\cdot\big[(\bm{p} + \bm{k}/2)\times\hat{\bm{e}}_\alpha\big] - (\hat{\bm{e}}_\alpha\cdot\hat{\bm{k}}) \bm{q}\cdot \big[(\bm{p}+\bm{k}/2)\times\hat{\bm{k}}\big] \Big\} \nonumber \\ 
    & \times \Bigg[\Bigg\{ |G_{\textrm{ret}}(0, \bm{p})|^2|G_{\textrm{ret}}(\omega_{\bm{k}},\bm{p}+\bm{k})|^2 \Bigg( \frac{1}{2}(\hat{\bm{e}}_i\cdot\bm{q}) (\hat{\bm{e}}_j\cdot\bm{p}) -\frac{1}{2}\hat{\bm{e}}_i\cdot(\bm{p} +\bm{k}) (\hat{\bm{e}}_j\cdot\bm{q}) \nonumber \\
    & + \hat{\bm{e}}_i\cdot(\bm{p}+\bm{k}) (\hat{\bm{e}}_j\cdot\bm{p})\bigg( \frac{\hbar^2\bm{p}\cdot\bm{q}}{m} G_{\textrm{ret}}(0, \bm{p}) + \frac{\hbar^2(\bm{p}+\bm{k})\cdot\bm{q}}{m} G_{\textrm{ret}}(\omega_{\bm{k}}, \bm{p}+\bm{k}) \bigg) \Bigg) \Bigg\} +\Bigg\{\omega_{\bm{k}} \longleftrightarrow -\omega_{\bm{k}}\Bigg\} \Bigg].
\end{align}
This expression is further simplified by approximating $(\bm{p}+\bm{k}/2)\approx \bm{p}$, $(\bm{p}+\bm{k})\approx\bm{p}$, and $G_\textrm{ret}(\omega_{\bm{k}}, \bm{p}+\bm{k}) \approx G_\textrm{ret}(0, \bm{p})$ \cite{mahan}. This approximation is ensured by the distribution factor $\frac{n_B(\omega_{\bm{k}})+n_F(\omega_{\bm{k}})}{\omega_{\bm{k}}}$ that exponentially decays for a large $|\bm{k}|$, effectively constraining $|\bm{k}| \ll \frac{Ck_BT}{\hbar c}$, where $C$ is an arbitrary constant. In a low-temperature regime $T \ll \frac{\hbar ck_F}{Ck_B}$ of our interest, this constraint indicates $|\bm{k}| \ll k_F \sim |\bm{p}|$, thus validating the stated approximation. Within this framework, we approximate  Eq.~\eqref{method:sigma1} as:
\begin{align} \label{method:sigma2} 
    \sigma^{\alpha}_{ij}(\bm{q}) = & \; \frac{i}{\pi}\frac{g^2e\hbar^6}{m^2M_{ph}}\int\frac{d^3p}{(2\pi)^3}\int\frac{d^3k}{(2\pi)^3} \frac{n_B(\omega_{\bm{k}})+n_F(\omega_{\bm{k}})}{2\omega_{\bm{k}}} |G_{\textrm{ret}}(0, \bm{p})|^4 \nonumber \\ 
    &\times \Big\{\bm{q}\cdot(\bm{p}\times\hat{\bm{e}}_\alpha) - (\hat{\bm{e}}_\alpha\cdot\hat{\bm{k}}) \bm{q}\cdot (\bm{p}\times\hat{\bm{k}}) \Big\} (q_ip_j - p_i q_j) \nonumber \\ 
    & + \frac{i}{\pi}\frac{g^2e\hbar^6}{m^2M_{ph}}\int\frac{d^3p}{(2\pi)^3}\int\frac{d^3k}{(2\pi)^3} \frac{n_B(\omega_{\bm{k}})+n_F(\omega_{\bm{k}})}{2\omega_{\bm{k}}} |G_{\textrm{ret}}(0, \bm{p})|^4 G_{\textrm{ret}}(0, \bm{p}) \nonumber \\ 
    &\times \Big\{\bm{q}\cdot(\bm{p}\times\hat{\bm{e}}_\alpha) - (\hat{\bm{e}}_\alpha\cdot\hat{\bm{k}}) \bm{q}\cdot (\bm{p}\times\hat{\bm{k}}) \Big\} p_i p_j \frac{4\hbar^2\bm{p}\cdot\bm{q}}{m}.
\end{align}
The requirement of the integrand being even in $\bm{k}$ and $\bm{p}$ leads to:
\begin{align} \label{method:pk_factors}
    & \Big\{\bm{q}\cdot(\bm{p}\times\hat{\bm{e}}_\alpha) - (\hat{\bm{e}}_\alpha\cdot\hat{\bm{k}}) \bm{q}\cdot (\bm{p}\times\hat{\bm{k}}) \Big\} (q_ip_j - p_i q_j) \rightarrow \Big\{ \bm{q}\cdot(\hat{\bm{e}}_j\times\hat{\bm{e}}_\alpha) q_ip_j^2 - \bm{q}\cdot(\hat{\bm{e}}_j\times\hat{\bm{e}}_\alpha) q_jp_i^2 \Big\} (1-\hat{k}_\alpha^2), \nonumber \\
    & \Big\{\bm{q}\cdot(\bm{p}\times\hat{\bm{e}}_\alpha) - (\hat{\bm{e}}_\alpha\cdot\hat{\bm{k}}) \bm{q}\cdot (\bm{p}\times\hat{\bm{k}}) \Big\} p_i p_j \rightarrow \bm{q}\cdot(\bm{p}\times\hat{\bm{e}}_\alpha) (1-\hat{k}_\alpha^2) p_i p_j.
\end{align}
Substituting Eq.~\eqref{method:pk_factors} into Eq.~\eqref{method:sigma2}, we obtain Eq.~\eqref{eq:sigmaDC_2}.

\subsection{Evaluation of the spin conductivity}

To obtain the explicit expression for $\sigma^{\alpha}_{ij}(\bm{q})$ presented in Eq.~\eqref{eq:sigmaDC_3}, we performed an analytic integration of Eq.~\eqref{eq:sigmaDC_2} by approximating the integrated as follows: (i) the electron propagator is represented as $G_{\textrm{ret}}(0, \bm{p})\approx (-\hbar v_Fp+i\hbar/2\tau)^{-1} \equiv G_{\textrm{ret}}(0, p)$, where $p\equiv|\bm{p}|-k_F$, (ii) the factors of $\bm{p}$ are replaced by $k_F\hat{\bm{p}}$, and (iii) the integral measure is represented as $\int d^3p \approx \int^\infty_{-\infty}dp k_F^2\int d\Omega_p$, where $\int d\Omega_p \equiv \int^{1}_{-1} d\cos{\theta_p} \int^{2\pi}_0d\phi_p $. These approximations are justified under the condition the condition $E_F\tau \gg \hbar$, ensuring that the integral over $\bm{p}$ sharply peeks around $|\bm{p}|\approx k_F$. Additionally, we represent the integral over $\bm{k}$ as $\int d^3k = \int^\infty_{0}dkk^2\int d\Omega_k$. Within this framework, we approximate $\sigma^\alpha_{ij}(\bm{q})$ in Eq.~\eqref{eq:sigmaDC_2} as:
\begin{align} \label{method:sigma3}
    \sigma^{\alpha}_{ij}(\bm{q}) = & \; \frac{i}{\pi} \frac{g^2e\hbar^6}{m^2M_{ph}} \frac{k_F^4q^2}{(2\pi)^6} \int^\infty_0dkk^2 \frac{n_B(\omega_{\bm{k}})+n_F(\omega_{\bm{k}})}{2\omega_{\bm{k}}} \int d\Omega_k (1-\hat{k}_\alpha^2) \nonumber \\ 
    & \times \int^\infty_{-\infty}dp |G_{\textrm{ret}}(0, p)|^4 \int d\Omega_{p} \Big[ \big\{(\hat{\bm{e}}_i \cdot \hat{\bm{q}})(\hat{\bm{e}}_j \cdot \hat{\bm{p}}) \hat{\bm{q}} \cdot( \hat{\bm{p}} \times \hat{\bm{e}}_\alpha) \big\} - \big\{i\leftrightarrow j\big\} \Big], \nonumber \\
    & + \frac{i}{\pi}\frac{g^2e\hbar^6}{m^2M_{ph}} \frac{4\hbar^2k_F^6q^2}{(2\pi)^6m} \int^\infty_0dkk^2 \frac{n_B(\omega_{\bm{k}})+n_F(\omega_{\bm{k}})}{2\omega_{\bm{k}}} \int d\Omega_k (1-\hat{k}_\alpha^2)  \nonumber \\
    & \times \int^\infty_{-\infty}dp |G_{\textrm{ret}}(0, p)|^4 G_{\textrm{ret}}(0, p)     \int d\Omega_{p}( \hat{\bm{e}}_i \cdot \hat{\bm{p}}) (\hat{\bm{e}}_j \cdot \hat{\bm{p}}) \big[\hat{\bm{q}} \cdot(\hat{\bm{p}} \times \hat{\bm{e}}_\alpha)\big].
\end{align}
The integrals for $\Omega_k$, $p$, and $\Omega_p$ can be conducted analytically, yielding
\begin{align} \label{method:integral1}
    \int d\Omega_k (1-\hat{k}_\alpha^2) & = \frac{8\pi}{3}, \nonumber \\
    \int^\infty_{-\infty}dp|G_{\textrm{ret}}(0, p)|^4 & = \frac{4\pi}{\hbar v_F(\hbar/\tau)^3},  \nonumber \\
    \int^\infty_{-\infty}dp |G_{\textrm{ret}}(0, p)|^4 G_{\textrm{ret}}(0, p) & = \frac{-6\pi i}{\hbar v_F(\hbar/\tau)^4}, 
\end{align}
and
\begin{align} \label{method:integral2}
    & \int d\Omega_{p} \Big[ \big\{(\hat{\bm{e}}_i \cdot \hat{\bm{q}})(\hat{\bm{e}}_j \cdot \hat{\bm{p}}) \hat{\bm{q}} \cdot( \hat{\bm{p}} \times \hat{\bm{e}}_\alpha) \big\} - \big\{i\leftrightarrow j\big\} \Big] \nonumber \\
    & = \frac{4\pi}{3} \Big[\big\{(\hat{\bm{e}}_i \cdot \hat{\bm{q}})\hat{\bm{q}} \cdot(\hat{\bm{e}}_j\times \hat{\bm{e}}_\alpha) \big\} - \big\{i\leftrightarrow j\big\} \Big], \nonumber \\
    & \int d\Omega_{p}( \hat{\bm{e}}_i \cdot \hat{\bm{p}}) (\hat{\bm{e}}_j \cdot \hat{\bm{p}}) \big[\hat{\bm{q}} \cdot(\hat{\bm{p}} \times \hat{\bm{e}}_\alpha)\big] (\hat{\bm{p}} \cdot \hat{\bm{q}}) \nonumber \\
    & = \frac{4\pi}{15}\Big[ \big\{(\hat{\bm{e}}_i \cdot \hat{\bm{q}})\big[\hat{\bm{q}} \cdot(\hat{\bm{e}}_j \times \hat{\bm{e}}_\alpha)\big]\big\} + \{i\leftrightarrow j\} \Big].
\end{align}
The remaining integral for $k$ is represented as:
\begin{align} \label{method:integral3}
    \int^\infty_0dkk^2 \frac{n_B(\omega_{\bm{k}})+n_F(\omega_{\bm{k}})}{2\omega_{\bm{k}}} = \frac{1}{k_B T}\bigg(\frac{k_BT}{\hbar c}\bigg)^3 I\bigg(\frac{E_g}{k_BT}\bigg),
\end{align}
where $I(y)$ is defined in Eq.~\eqref{eq:I_1}. Substituting Eqs.~\eqref{method:integral1} to \eqref{method:integral3} into Eq.~\eqref{method:sigma3}, we obtain Eq.~\eqref{eq:sigmaDC_3}.


\subsection{Numerical integration method} \label{sec:numerical_integration}

To produce the results depicted in Figs.~4 and 6, we performed numerical integration of the $\sigma_{ij}^\alpha(\bm{q})$ formula given in Eq.~\eqref{eq:sigmaDC_1} using a Riemann sum approach. The discretization of the electron wave vector $\bm{p} = (p_x, p_y, p_z)$ spanned the Fermi surface region, defined as $\sqrt{k_F^2 - 9.95 \cdot (\hbar/2\tau)} < |\bm{p}| < \sqrt{k_F^2 + 9.95 \cdot (\hbar/2\tau)}$. At the boundaries of this region, we ensured that the electron spectral function, $-\frac{1}{\pi}\textrm{Im}[G_\textrm{ret}]$, descended below 1\% of its maximum value occurring at $|\bm{p}| = k_F$. Consequently, contributions from outside this region were deemed negligible. Additionally, we established a mesh for the phonon wave vector $\bm{k}=(k_x, k_y, k_z)$ within the range $|\bm{k}|\leq k_\textrm{cutoff}=10\frac{k_BT}{\hbar c}$. At the boundaries of this region, we ensured that the distribution factor ${n_B(\omega_{\bm{k}})+n_F(\omega_{\bm{k}})}$ fell below a few percent of its maximum value occurring at $|\bm{k}|=0$ (the ratio was maintained at 1\% for $E_g=0$ and less than 10\% for $E_g \approx 10k_BT$). Consequently, contributions from outside this region were also considered negligible. For the computation, we used a grid size of $(51\times51\times51)$ for both $(p_x, p_y, p_z)$ and $(k_x, k_y, k_z)$ within the specified regions for numerical integration. Our analyses confirmed that the numerical integration results were well-converged and reliable for our intended purposes.

\section{Data availability} 
The data for the results presented in the paper and Supplementary Information are available from the corresponding author upon reasonable request. 

\section{Acknowledgements}
We would like to thank Jung Hoon Han, Beom Hyun Kim, Changyoung Kim, Ki-Seok Kim, Se Kwon Kim, Steve Kivelson, Hyun-Woo Lee, and S. Raghu for sharing their insights. K.-M.K. was supported by the Institute for Basic Science in the Republic of Korea through the project IBS-R024-D1. S.B.C. was supported by the National Research Foundation of Korea (NRF) grants funded by the Korea government (MSIT) (NRF-2023R1A2C1006144, NRF-2020R1A2C1007554, and NRF-2018R1A6A1A06024977).

\section{Author information}

{\it Contributions:} K.-M.K. performed numerical calculation. 
K.-M.K. and S.B.C. 
designed the project, derived formulae and 
wrote the manuscript. K.-M.K. and S.B.C. hereby confirm that they have read and approved the manuscript.  

{\it Corresponding authors:} Correspondence to either Kyoung-Min Kim (kmkim@ibs.re.kr) or Suk Bum Chung (sbchung0@uos.ac.kr).

\section{Inclusion \& Ethics statement}
{\it Competing interests:} The authors declare no competing interests.

\end{document}